\documentclass[manuscript,screen]{acmart}
\usepackage[utf8]{inputenc}
\usepackage{hyperref}
\usepackage{booktabs}
\usepackage[ruled,linesnumbered]{algorithm2e}
\usepackage{subcaption}
\usepackage{multicol}
\usepackage{todonotes}
\usepackage{optidef}

\begin{document}
\newcommand{\csecunige}{Università degli Studi di Genova}
\newcommand{\pralab}{Università degli studi di Cagliari}
\newcommand{\pluribus}{Pluribus One}
\newcommand{\fireye}{FireEye, Inc.}

\newcommand{\bytestrings}{\{0, \dots , 255\}^*}
\newcommand{\programs}{\mathcal{Z}}
\newcommand{\faspace}{\mathcal{T}}
\newcommand{\query}{\mathcal{Q}}
\newcommand{\size}{\mathcal{C}}
\newcommand{\algoname}{{RAMEN}\xspace} 

\newcommand{\ember}{{EMBER}\xspace}
\newcommand{\embersize}{{1.1 M}\xspace}
\newcommand{\largedatasetsize}{{16.3M}\xspace}
\newcommand{\testdataset}{104\xspace}
\newcommand{\inember}{37\xspace}
\newcommand{\packed}{46\xspace}

\newcommand{\diff}[2]{\frac{\partial #1}{\partial #2}}
\newcommand{\vct}[1]{\ensuremath{\boldsymbol{#1}}}
\newcommand{\mat}[1]{\ensuremath{\mathbf{#1}}}
\newcommand{\set}[1]{\ensuremath{\mathcal{#1}}}
\newcommand{\con}[1]{\ensuremath{\mathsf{#1}}}
\newcommand{\tsum}{\ensuremath{\textstyle \sum}}
\newcommand{\T}{\ensuremath{\top}}
\newcommand{\mycomment}[1]{\textcolor{red}{#1}}
\newcommand{\ind}[1]{\ensuremath{\mathbbm 1_{#1}}}
\newcommand{\argmax}{\operatornamewithlimits{\arg\,\max}}
\newcommand{\erf}{\text{erf}}
\newcommand{\sign}{\text{sign}}
\newcommand{\argmin}{\operatornamewithlimits{\arg\,\min}}
\newcommand{\bmat}[1]{\begin{bmatrix}#1\end{bmatrix}}
\newcommand{\myparagraph}[1]{\vspace*{0.05cm} \noindent \textbf{#1}}
\newcommand{\ie}{i.e.\xspace}
\newcommand{\eg}{e.g.\xspace}
\newcommand{\etc}{etc.\xspace}
\newcommand{\aka}{a.k.a.\xspace}
\newcommand{\wrt}{w.r.t.\xspace}
\newcommand{\SoA}{state of the art\xspace}


\setcopyright{acmcopyright}

\acmDOI{TO EDIT}

\acmISBN{TO EDIT}

\acmConference[]{ACM TO EDIT conference}{July 2019}{EDIT}

\acmYear{2020}

\copyrightyear{2020}




\title[Adversarial {EXE}mples: Practical Attacks on Machine Learning for Windows Malware Detection]{Adversarial {EXE}mples: A Survey and Experimental Evaluation of Practical Attacks on Machine Learning for Windows Malware Detection}
\author{Luca Demetrio}
\affiliation{%
   \institution{\pralab}
   \city{Cagliari}
   \country{ITA}}
\email{luca.demetrio@dibris.unige.it}

\author{Scott E. Coull}
\affiliation{%
   \institution{\fireye}
}
\email{scott.coull@fireye.com}

\author{Battista Biggio}
\affiliation{%
   \institution{\pralab}
   \city{Cagliari}
   \country{ITA}}
\affiliation{%
   \institution{\pluribus}
   \city{Cagliari}
   \country{ITA}}
\email{battista.biggio@unica.it}

\author{Giovanni Lagorio}
\affiliation{%
   \institution{\csecunige}
   \city{Genova}
   \country{ITA}}
\email{giovanni.lagorio@unige.it}

\author{Alessandro Armando}
\affiliation{%
   \institution{\csecunige}
   \city{Genova}
   \country{ITA}}
\email{alessandro.armando@unige.it}

\author{Fabio Roli}
\affiliation{%
   \institution{Università degli Studi di Cagliari}
   \city{Cagliari}
   \country{ITA}}
\affiliation{%
   \institution{Pluribus One}
   \city{Cagliari}
   \country{ITA}}
\email{fabio.roli@unica.it}

\renewcommand{\shortauthors}{Demetrio et al.}

\begin{abstract}
Recent work has shown that adversarial Windows malware samples - referred to as adversarial \emph{EXE}mples in this paper - can bypass machine learning-based detection relying on static code analysis by perturbing relatively few input bytes.
To preserve malicious functionality, previous attacks either add bytes to existing non-functional areas of the file, potentially limiting their effectiveness, or require running computationally-demanding validation steps to discard malware variants that do not correctly execute in sandbox environments.
In this work, we overcome these limitations by developing a unifying framework that does not only encompass and generalize previous attacks against machine-learning models, but also includes three novel attacks based on practical, functionality-preserving manipulations to the Windows Portable Executable (PE) file format. These attacks, named \textit{Full DOS}, \textit{Extend} and \textit{Shift}, inject the adversarial payload by respectively manipulating the DOS header, extending it, and shifting the content of the first section. 
Our experimental results show that these attacks outperform existing ones in both white-box and black-box scenarios, achieving a better trade-off in terms of evasion rate and size of the injected payload, while also enabling evasion of models that have been shown to be robust to previous attacks. To facilitate reproducibility of our findings, we open source our framework and all the corresponding attack implementations as part of the \texttt{secml-malware} Python library. 
We conclude this work by discussing the limitations of current machine learning-based malware detectors, along with potential mitigation strategies based on embedding domain knowledge coming from subject-matter experts directly into the learning process.
\end{abstract}


\maketitle

\section{Introduction}

Machine learning (ML) has become an important aspect of modern cybersecurity due to its ability to detect new threats far earlier than signature-based defenses.
While many cybersecurity companies use machine learning models\footnote{\url{https://www.sophos.com/products/intercept-x/tech-specs.aspx}}\footnote{\url{https://www.fireeye.com/blog/products-and-services/2018/07/malwareguard-fireeye-machine-learning-model-to-detect-and-prevent-malware.html}}\footnote{\url{https://www.kaspersky.com/enterprise-security/wiki-section/products/machine-learning-in-cybersecurity}}\footnote{\url{https://www.avast.com/technology/ai-and-machine-learning}} in their respective product offerings, creating and maintaining these models often represents a significant cost in terms of expertise and labor in developing useful features to train on, particularly when we consider that each new file type may require a completely different set of features to provide meaningful classification.
With this in mind, researchers have recently proposed end-to-end deep learning models that operate directly on the raw bytes of the input files and automatically learn useful feature representations during training, without external knowledge from subject-matter experts.
Several byte-based malware detection models for Windows PE files, for example, have demonstrated efficacy that is competitive with traditional ML models~\cite{raff2018malware, coull2019activation} (Sect.~\ref{sec:dl_malware}).

While the use of end-to-end deep learning makes it easy to create new models for a variety of file types by simply exploiting the vast number of labeled samples available to such organizations, it also opens up the 
possibility of attacking these models using \emph{adversarial evasion} techniques popularized in the image classification space. 
In particular, recent work has shown how an attacker can create what we call here \emph{adversarial EXEmples}, \ie,  Windows malware samples carefully perturbed to evade learning-based detection while preserving malicious functionality~\cite{kolosnjaji2018adversarial, demetrio2019explaining, demetrio2020efficient, anderson2017evading, castro2019aimed, kreuk2018deceiving, suciu2019exploring, sharif2019optimization}.

Unlike adversarial evasion attacks in other problem areas, such as image classification, manipulating malware while simultaneously preserving its malicious payload can be difficult to accomplish.
In particular, each perturbation made to the input bytes during the attack process may lead to changes in the underlying syntax, semantics, or structure that could prevent the binary from executing its intended goal.
To address this problem, the attacker can take one of two approaches: apply invasive perturbations and use dynamic analysis methods (\eg emulation) to ensure that functionality of the binary is not compromised~\cite{castro2019aimed,song2020automatic}, or focus the perturbation on areas of the file that do not impact functionality~\cite{kreuk2018deceiving, suciu2019exploring, demetrio2020efficient, demetrio2019explaining, kolosnjaji2018adversarial} (\eg appending bytes).
Naturally, this leads to a trade-off between strong yet time-consuming attacks on one extreme, and weaker but more computationally-efficient attacks on the other.

In this work, we overcome these limitations by proposing a unifying framework, called \textbf{\algoname} (Sect.~\ref{sec:aml_malware}), built on top of a family of \emph{practical} manipulations to the Windows Portable Executable (PE) file format that can alter the structure of the input malware without compromising its semantics. 
Our framework encompasses and generalizes previously-proposed attacks against learning-based Windows malware detectors based on static code analysis, including both white-box attacks that exploit full knowledge of the target algorithm, and black-box attacks that only require query access to it.
The practical, functionality-preserving manipulations defined in our framework are not limited to perturbing bytes at the end of malware programs and do not require computationally-demanding validation steps during the attack optimization, thereby overcoming the limitations of existing attacks.
In particular, we encode three novel practical manipulations that exploit the ambiguity in the specifications of the Windows PE file format: \emph{Full DOS} that edit all the available bytes inside the DOS header; \emph{Extend}, which enlarges the DOS header, thus enabling manipulation of these extra DOS bytes; and \emph{Shift}, which shifts the content of the first section, carving additional space for the adversarial payload.

Our experimental results (Sect.~\ref{sec:experiments}) show that these attacks outperform existing ones in both white-box and black-box attack scenarios against different machine-learning models, deep network architectures, activation functions (\ie linear vs. non-linear models), and training regimes. 
In particular, our \emph{Extend} and \emph{Shift} attacks enable evading some models that are not affected by previously-proposed attacks, while generally achieving a better trade-off in terms of evasion rate and size of the injected payload; they create fully-functional, evasive malware by perturbing roughly 2\% of the input bytes against most of the considered classifiers.

An additional finding from our experimental analysis is that, while dataset size and activation functions do not seem to play a significant role in improving adversarial robustness, model architecture does, at least to some extent, with all attacks working well against Raff et al.'s MalConv classifier~\cite{raff2018malware} and only content-shifting attacks working well against Coull et al.'s classifier~\cite{coull2019activation}, possibly due to the importance of spatial locality in its design.
This identifies a promising line of research towards strengthening models against adversarial attacks through inclusion of additional structure in the training process. 
We conclude the paper by discussing related work (Sect.~\ref{sec:related}) along with the limitations of our methodology (Sect.~\ref{sec:limitations}), and promising research directions to improve robustness of learning-based Windows malware detectors against adversarial attacks (Sect.~\ref{sec:conclusions}).
Besides considering network architectures that exploit spatial locality, we discuss other potential strategies to
embed external domain knowledge directly into the learning process (e.g., via suitable constraints and loss functions) with the goal of learning more meaningful and robust representations from data~\cite{melacci2020domain}. 
We believe that this novel learning paradigm may help significantly improve adversarial robustness of such models, while at the same time exploiting knowledge from domain experts in an efficient manner.

To summarize, we highlight our contributions below.
\begin{itemize}
     \item We propose \algoname, a general framework for expressing white-box and black-box adversarial attacks on learning-based Windows malware detectors based on static code analysis.
    \item We propose three novel attacks based on practical, functionality-preserving manipulations, named \emph{Full DOS}, \emph{Extend} and \emph{Shift}, which improve the trade-off between the probability of evasion and the amount of manipulated bytes in both white-box and black-box attack settings.
     \item We release the implementations of all the aforementioned white-box and black-box attacks encompassed by our framework (including previous attacks, \emph{Extend} and \emph{Shift}), as an open-source project available at \url{https://github.com/zangobot/secml_malware}.
     \item We identify promising future research directions towards improving adversarial robustness of learning-based Windows malware detectors leveraging static code analysis.
 \end{itemize}

\section{Background}
\label{sec:dl_malware}
Before diving into the details of our proposed attack framework, we first provide some necessary background on the Windows Portable Executable (PE) file format (Section \ref{sec:windows_pe})  and the malware classifiers that we will examine in our experiments (Section \ref{sec:learn-malic-from}).

\subsection{Executable File Format}
\label{sec:windows_pe}
\begin{figure}
    \centering
    \includegraphics[width=0.6\textwidth]{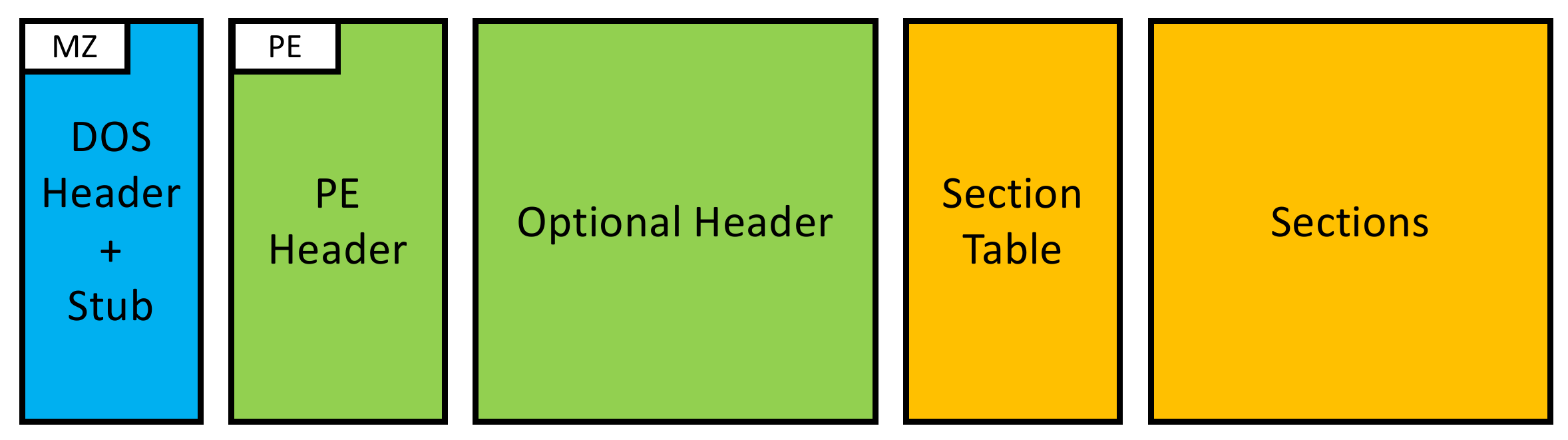}
    \caption[]{The Windows PE file format.
    Each colored section describes a particular characteristic of the program.
    }
    \label{fig:pe_format}
\end{figure}

The \emph{Windows Portable Executable} (PE)\footnote{\url{https://docs.microsoft.com/en-us/windows/win32/debug/pe-format}} format specifies how executable programs are stored as a file on disk.
The OS loader parses this structure and maps the code and data into memory, following the directives specified by the header of the file.
We show the components of the format in Figure \ref{fig:pe_format}.

\myparagraph{DOS Header and Stub.}
The DOS header contains metadata for loading the executable inside a DOS environment, while the DOS Stub is made up of few instructions that will print ``\emph{This program cannot be run in DOS mode}'' if executed inside a DOS environment.
These two components have been kept to maintain compatibility with older Microsoft operating systems.
From the perspective of a modern application, the only relevant locations contained inside the DOS Header are (i) the magic number \verb|MZ|, a two-byte signature of the file, and (ii) the four-byte integer at offset \verb|0x3c|, that works as a pointer to the actual header.
If one of these two values is altered for some reason, the program is considered corrupted, and it will not be executed by the OS.

\myparagraph{PE Header.}
It is the real header of the program, and it contains the magic number \verb|PE| and the characteristics of the executable, such as the target architecture that can run the program, the size of the header and the attributes of the file.

\myparagraph{Optional Header.}
Not optional for executables and DLLs, it contains the information needed by the OS for loading the binary into memory.
Among these fields, the Optional Header specifies the (i)\emph{file alignment}, that acts as a constraint on the structure of the executable since each section of the program must start at an offset multiple to that field, and the (ii) \emph{size of headers} that specifies the amount of bytes that are reserved to all the headers of the programs, and it must be a multiple of the \emph{file alignment}.
Lastly, the optional header contains offsets that point to useful structures, like the Import Table needed by the OS for resolving dependencies, the Export Table to find functions that can be referenced by other programs, and more.

\myparagraph{Section Table.}
It is a list of entries that indicates the characteristics of each section of the program.
Each section entry has a name, an offset to the location inside the binary, a virtual address where the content should be mapped in memory, and the characteristics of such content (\ie is read-only, write-only, or it is executable, and more).

\myparagraph{Sections.}
These are contiguous chunks of bytes, loaded in memory by the loader after while parsing the Section Table.
To name a few, there is the code of the program (\emph{.text} section), initialized data (\emph{.data}), read-only constants (\emph{.rdata}), and counting.
To maintain the alignment specified inside the Optional Header, these sections might be zero-padded to match the format constraint.
It is clear that, even without executing the program contained inside a file, it is possible to infer some useful information from its headers, imports, exports, and sections.

\subsection{Malware Classifiers}
\label{sec:learn-malic-from}
Here, we describe two recent byte-based convolutional neural network models for malware detection.
Both take as input the raw bytes from the Windows PE file on disk, use an embedding layer to encode the bytes into a higher-dimensional space, and then apply one or more convolutional layers to learn relevant features that are fed to a fully-connected layer for classification with a sigmoid function.
While they share common design concepts, they differ in their overarching architecture and, as we will see, this difference is the key to their respective robustness to the various adversarial evasion attacks described in this paper.
In addition to these two deep learning models, we also consider a traditional ML model using gradient boosting decision trees on hand-engineered features, which we use as a baseline for purposes of comparison and to evaluate attack transferability from byte-based models to models with semantically-rich features.
In general, both neural networks and standard ML algorithms, can not really work in an end-to-end way, since bytes are categorical data that do not possess a defined metric.
In practice, the pipeline for predicting the maliciousness from an input program is the same, as shown in Figure \ref{fig:model_scheme}.

\myparagraph{MalConv.} This model, proposed by Raff et al.~\cite{raff2018malware}, is a convolutional neural network that combines an 8-dimensional, learnable embedding layer with a 1-dimensional gated convolution.
The embedding layer acts as a non-differentiable feature mapping, as it maps each input byte (treated as a categorical value) onto a specific point in the embedded space. The goal of this step is to learn to represent bytes that exhibit a semantically-similar behavior as closer points in this space, thus obtaining a meaningful distance measure between bytes.  
The convolutional layer iterates over non-overlapping windows of 500 bytes each, with a total of 128 convolutional filters.
A global max pooling is applied to the gated outputs of the convolutional layer to select the 128 largest-activation features, without considering the structure or locality of those features within the binary. The corresponding values are then used as input to a fully-connected layer for classification.
While the original MalConv model considers a maximum input file size of 2MB, the model used in our experiments is that provided by Anderson et al.~\cite{anderson2018ember}, trained on the EMBER dataset with a maximum input file size of 1MB.
Files exceeding the maximum allowable size are truncated, while shorter files are padded using a special padding token separate from the standard bytes in the file (\ie, resulting in 257 unique tokens).

\myparagraph{DNN with Linear (DNN-Lin) and ReLU (DNN-ReLU) activations.} Jeffrey Johns\footnote{\url{https://www.camlis.org/2017/jeffreyjohns}} and Coull et al.~\cite{coull2019activation} proposed a deep convolutional neural network that combines a 10-dimensional, learnable embedding layer with a series of five interleaved convolutional and max-pooling layers arranged hierarchically so that the original input size is reduced by one quarter (1/4) after each layer.
The outputs of the final convolutional layer are globally pooled to create a fixed-length feature vector that is then provided as input to a fully-connected layer for the final classification.
Since the convolutional layers are hierarchically arranged, locality information among the learned features is preserved and compressed as it flows upwards towards the final classification layer.
The maximum length of this model is 100KB to account for the deep architecture and, as done by MalConv, files exceeding this length are truncated, while shorter files are padded with a special padding token.
Several variations of this architecture are evaluated in this paper, including examining performance with both linear and Rectified Linear Unit (ReLU) activations for the convolutional layers, as well as performance when trained using the EMBER dataset and a proprietary dataset containing more than 10x the number of training samples.
An analysis of the model by Coull et al.~\cite{coull2019activation} demonstrates how the network attributes importance to meaningful features inside the binary, such as the name of sections, the presence of the checksum, and other structures.

\myparagraph{Gradient Boosting Decision Tree (GBDT).}  A gradient-boosted tree ensemble model trained provided as part of the EMBER open-source dataset by Anderson et al~\cite{anderson2018ember}.  
The model uses a set of 2,381 hand-engineered features derived from static analysis of the binary using the LIEF PE parsing library, including imports, byte-entropy histograms, header properties, and sections, which generally represent the current {\SoA} in traditional ML-based malware detection.
Given its use of a diverse set of semantically-meaningful static features, it provides an excellent baseline to compare the two above byte-based models against, and help demonstrate the gap between the features learned by byte-based neural networks and those created by subject-matter experts.

The MalConv architecture has been extensively studied by previous work and a wide variety of adversarial attacks have shown great success against the model~\cite{kolosnjaji2018adversarial, demetrio2019explaining, kreuk2018deceiving, suciu2019exploring, demetrio2020efficient, sharif2019optimization}.
As pointed out by Suciu et al.~\cite{suciu2019exploring}, the lack of robustness in this model may be strongly tied to its weak notions of spatial locality among the learned features -- meaning that the location of the injected adversarial noise does not matter as long as the activation on the noise bytes overwhelms those from the actual binary.
By contrast, the deep convolutional net proposed by Johns and Coull et al.~\cite{coull2019activation} enforces spatial locality among features, which means both the location and magnitude of adversarial noise play a role in the success of evasion attacks.
Even the GBDT model has been shown to be vulnerable to evasion attacks~\cite{castro2019aimed, demetrio2020efficient, sharif2019optimization}, albeit with more advanced and computationally-intensive attacks.
While each of these models has been previously evaluated in an ad-hoc manner, we are the first to treat attacks on machine-learning models in a holistic manner using a single unifying framework, and in doing so we uncover two new attack methods that apply to both MalConv and the Coull et al.'s model despite the unique architectural differences between the two.

\section{Adversarial \textit{EXE}mples: Practical Attacks on Windows Malware Detectors}
\label{sec:aml_malware}

In this section, we introduce \algoname, our framework for the optimization of adversarial malware with practical manipulations. 
We first formalize the problem of optimizing adversarial examples with application-specific manipulation constraints under a unifying attack framework (Section~\ref{sect:framework}), inspired from previous work in~\cite{pierazzi2019intriguing,biggio14-tkde,huang11}, and discuss how to implement gradient-based (white-box) and gradient-free (black-box) attacks within this framework (Sections~\ref{sect:grad-based}-\ref{sect:grad-free}).
We then discuss the novel, functionality-preserving manipulation strategies identified in this work to manipulate Windows malware files (\ie, \emph{Full DOS}, \emph{Extend} and \emph{Shift}), along with the other previously-proposed practical manipulations that are encompassed by our framework
(Section~\ref{sec:practical_manipulations}).
We conclude by discussing how to implement practical white-box (Section~\ref{subsec:wbox_implement}) and black-box (Section~\ref{subsec:bbox_implement}) attacks on Windows malware detectors based on the aforementioned manipulation strategies, detailing the implementation of our novel attacks and how to recast previously-proposed ones into our framework.

\subsection{Attack Framework}
\label{sect:framework}

\begin{figure}
    \centering
    \includegraphics[width=0.9\linewidth]{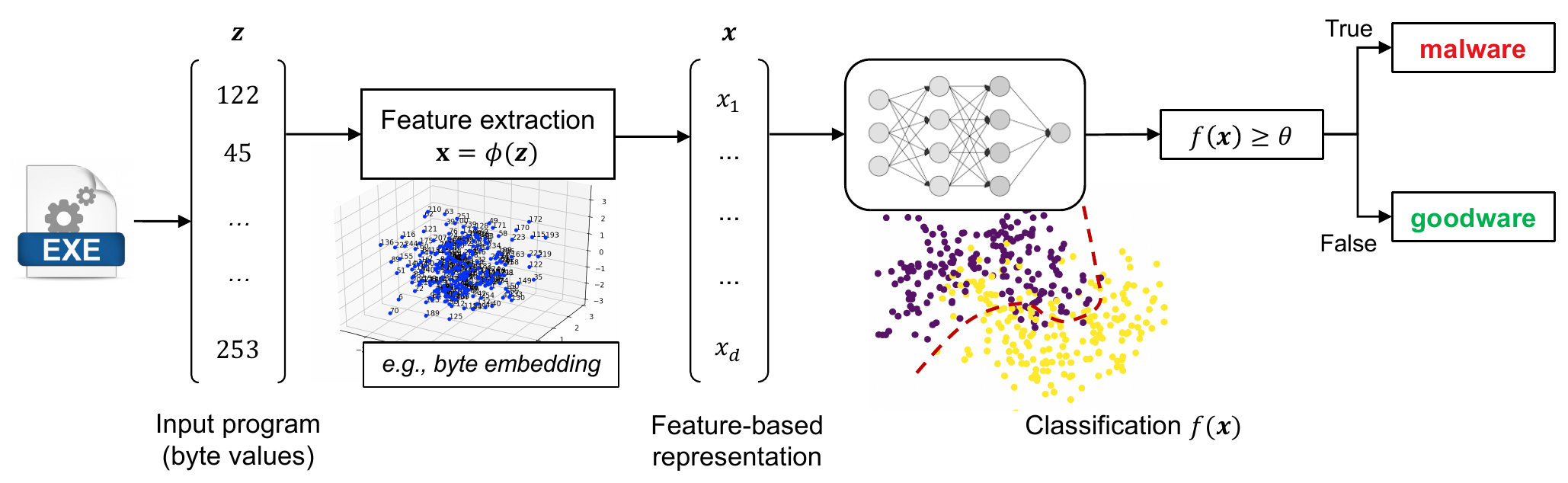}
    \caption{Conceptual representation of malware detectors based on machine learning.}
    \label{fig:model_scheme}
\end{figure}

We consider here machine-learning models that take a program as input and aim to classify it either as legitimate or malicious, as shown in Figure~\ref{fig:model_scheme}.
Since code is written as arbitrarily-long strings of bytes, we define the set of all possible functioning programs in the input space as $\programs \subset \bytestrings$.
Most classifiers process data through a \emph{feature extraction} step, \eg, either embedding the input bytes on a representation space which reflects their semantic similarity, or using handcrafted feature values like byte histograms, n-grams, and other meaningful statistics extracted from the input file structure and content.
We encode this step as $\phi : \programs \rightarrow \mathcal{X}$, being $\mathcal{X} \subseteq \mathbb{R}^d$ a $d$-dimensional vector space (\ie, the feature space). The prediction function is denoted with $f : \mathcal{X} \rightarrow \mathbb{R}$. 
We assume here that, without loss of generality,
this function outputs a continuous value representing the probability of the input sample belonging to the malware class. 
Then, we compute $f(\vct x) \geq \theta$, where $\theta$ is the detection threshold, to obtain a final decision in the label space $\mathcal{Y} = \{-1, +1\}$, being $-1$ and $+1$ the class labels for legitimate and malicious samples, respectively.

Under this setting, the attacker aims to craft adversarial malware by perturbing each malicious input program to achieve evasion with high confidence.
To this end, the attacker is required to apply \emph{practical manipulations}, \ie, transformations that alter the representation of the input program without disrupting its original behavior, by exploiting the redundancies offered by the executable file format.
We encode these functionality-preserving manipulations as a function $h : \programs \times \faspace \rightarrow \programs$, whose output is a functioning program with the same behavior as the input, but with a different representation.
The function $h$ takes as input a program $\vct z \in \programs$ and a vector $\vct t \in \faspace$, representing the parameters of the applied transformation.
For example, the injection of $K$ padding bytes at the end of the file, which is a functionality-preserving manipulation~\cite{kolosnjaji2018adversarial,demetrio2020efficient}, can be encoded in our framework as $\vct t=(t_1, \ldots, t_K)$, being $t_i$ the value of the $i^{\rm th}$ padding byte. Accordingly, the function $h(\vct z, \vct t)$ will return a manipulated program $\vct z^\prime$ consisting of the input program $\vct z$ with the padding bytes $\vct t$ appended at the end.

We are now in the position to present \textit{\algoname}, a general framework that reduces the problem of computing adversarial malware examples to optimization problems of the form:
\begin{mini}|l|
    {\vct t \in \faspace}{F(\vct t) = L ( f ( \phi (h(\vct z, \vct t)), y)}{}{} \, .
    \label{eq:ramen}
\end{mini}
where $L : \mathbb{R} \times \mathcal{Y} \rightarrow \mathbb{R}$ is a \emph{loss function} that measures how likely an input sample is classified as malware, by comparing the output of the prediction $f(\phi(\vct z))$ on a malicious input sample $\vct z$ against the class label $y=-1$ of benign samples.
By minimizing this loss function, the attacker aims to reduce the probability of the modified program being recognized as malware, \ie, increases the probability of evasion, while retaining malicious functionality.
We discuss the two main strategies for solving this optimization problem in the following, which include gradient-based and gradient-free attacks.

\subsection{Gradient-based (White-box) Attacks}
\label{sect:grad-based}

\begin{figure}[t]
    \centering
    \includegraphics[width=\textwidth]{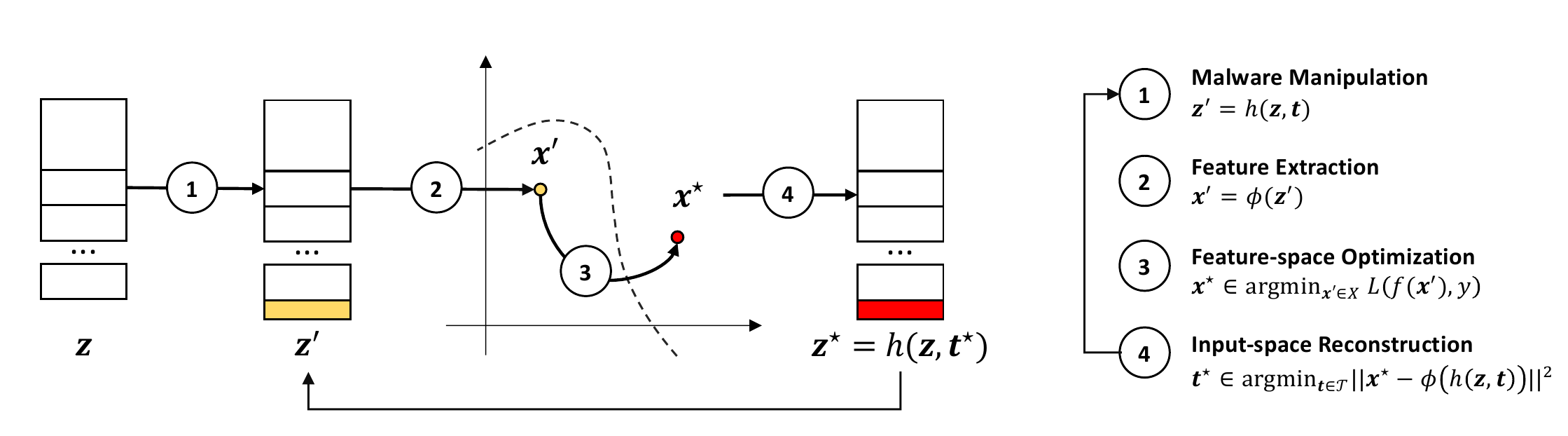}
    \caption{The steps performed by \algoname to optimize gradient-based (white-box) attacks, as also detailed in Algorithm~\ref{algo:ramen_wb_general}.}
    \label{fig:ramen_scheme}
\end{figure}

The aforementioned optimization problem can be solved, at least in principle, using gradient-based approaches, similarly to attacks staged against machine-learning algorithms for image classification~\cite{biggio13-ecml,szegedy14-iclr}. 
This implicitly assumes that the attacker has white-box access to the target model to compute the gradient of the loss function $L$ (as it requires knowledge of the model's internal parameters).
However, when optimizing adversarial Windows malware, the loss gradient can not be typically used to directly update the parameter vector $\vct t$ (\eg, the padding bytes), due to the non-differentiability of the inner feature-mapping function $\phi$.
For example, let us consider the case in which the input program bytes are embedded in a vector space, as done by MalConv, and our manipulation strategy $h(\vct z, \vct t)$ amounts to injecting padding bytes.
In this scenario, it is not possible to compute the loss gradient with respect to the bytes in $\vct t$, as the embedding step performed by MalConv is not differentiable (in particular, bytes are treated as categorical variables, like words, rather than as numerical values).
To overcome this issue, most of the gradient-based attacks on Windows malware proposed so far have considered optimizing the attack by performing gradient descent in the feature space, while iteratively trying to reconstruct the corresponding adversarial malware example in the input space, using different strategies.
In the following, we formalize this process in the context of our framework, according to the steps detailed in Algorithm~\ref{algo:ramen_wb_general} and  Figure~\ref{fig:ramen_scheme}.

\begin{algorithm}[t]
    \SetKwInOut{Input}{Input}
    \SetKwInOut{Output}{Output}
    \KwData{$\vct z$, the initial malware sample; $N$, total number of iterations; $y$, the target class label; $f$, the target model.}
    \KwResult{$\vct z^\star$, the adversarial EXEmple.}
    $\vct t^{(0)} \in \faspace$\\
    \textbf{for} $ i $ \textbf{in} $ [0, N-1]$\\
    \Indp
        $\vct z^\prime \leftarrow h(\vct z, \vct t^{(i)})$\label{line:ramen_1} \# \emph{malware manipulation}\\
        $\vct x^\prime \leftarrow \phi(\vct z^\prime)$ \label{line:ramen_2} \# \emph{feature extraction}\\
        $\vct x^\star \leftarrow \argmin_{\vct x^\prime \in \set X} L(f(\vct x^\prime), y)$ \# \emph{feature-space optimization}\label{line:ramen_3}\\
        $\vct t^{(i+1)} \leftarrow \argmin_{\vct t \in \faspace} \| \vct x^\star - \phi(h(\vct z, \vct t^{(i)})) \|^2$ \# \emph{input-space reconstruction} \label{line:ramen_4}\\
    \Indm
    $\vct t^\star \leftarrow \vct t^{(N)}$\\
    $\vct z^\star \leftarrow h(\vct z, \vct t^\star)$\\
    return $\vct z^\star$\label{line:ramen_return}
    \caption{Gradient-based (White-box) Attacks for Optimizing Adversarial Malware EXEmples in \algoname.}
    \label{algo:ramen_wb_general}
\end{algorithm}

\myparagraph{Malware Manipulation.} The first step consists of applying an initial perturbation $h(\cdot, \vct t)$ to the input malware program $\vct z$, obtaining a modified program $\vct z^\prime = h(\vct z, \vct t)$ (line~\ref{line:ramen_1}).
While we discuss in detail the practical manipulations that are encompassed by our framework in Section~\ref{sec:practical_manipulations}, the reader may consider here the injection of $K$ randomly-chosen padding bytes at the end of the file (\ie, the yellow region in Figure~\ref{fig:ramen_scheme}) as an exemplary case.

\myparagraph{Feature Extraction.} The second step (line~\ref{line:ramen_2}) amounts to encoding the manipulated program $\vct z^\prime$ in the feature space as $\vct x^\prime = \phi(\vct z^\prime)$ (\eg, through byte embedding in MalConv).

\myparagraph{Feature-space Optimization.} The third step (line~\ref{line:ramen_3}) consists of modifying the feature-based representation $\vct x^\prime$ of the manipulated input sample $\vct z^\prime$ to minimize the loss function $L$, using gradient-based updates. It can be formalized as:
\begin{equation}
    \vct x^\star \in \argmin_{\vct x^\prime \in \set X} L(f(\vct x^\prime), y) \, .
    \label{eq:feat_space_search}
\end{equation}
The solution is obtained by iteratively updating the feature-based representation as $\vct x^\prime \leftarrow \vct x^\prime - \eta \nabla_{\vct x} L(f(\vct x^\prime), y)$, being $\eta$ the gradient step size, until convergence or a maximum number of iterations is reached. As we will discuss in the remainder of this work, some attacks just consider a single gradient update in this step. It is worth remarking here that the feature-based representation $\vct x^\star$ obtained after this step may not necessarily correspond to any input sample $\vct z \in \set Z$. For this reason, it is necessary to use proper reconstruction strategies to ensure not only that the reconstructed sample is a valid program, but that it also preserves the intended functionality of the initial malware program $\vct z$.

\myparagraph{Input-space Reconstruction.} The final step (line~\ref{line:ramen_4}) optimizes the parameters $\vct t^\star$ to generate a functionality-preserving malware sample $\vct z^\star$ whose feature-based representation is as close as possible to the desired $\vct x^\star$. This can be expressed as a reconstruction problem in the form of a minimization:
\begin{equation}
    \vct t^\star \in \argmin_{\vct t \in \faspace} \| \vct x^\star - \phi(h(\vct z^\prime, \vct t)) \|^2 \, .
    \label{eq:reconstruction}
\end{equation}
For attacks that only consider a single gradient update in the previous step, the reconstruction step is similarly performed by finding the transformation $\vct t$ that is best aligned with the loss gradient in feature space, \ie, $\vct t^\star \in \argmax_{\vct t \in \set T} \nabla_x L(f(\vct x^\prime), y)^\T \left (\vct x^\prime-\phi(h(\vct z^\prime, \vct t)) \right )$~\cite{sharif2019optimization}.
Note that these gradient-based attacks are only convenient if the reconstruction step can be computed efficiently; otherwise, gradient-free attacks may be preferable. 
For instance, let us assume that the input bytes are embedded as points in a bi-dimensional vector space; \eg, bytes $0$ and $1$ are encoded as $0 \mapsto (1, 2)$ and $1 \mapsto (1, 5)$, respectively. If we consider the injection of padding bytes (\ie, the yellow region in Figure~\ref{fig:ramen_scheme}), the feature-based representations of such bytes will be shifted by the attack along the gradient direction in the embedding space, and may result in feature values that do not correspond to any byte; \eg, the initial embedding of byte $0$ may be modified from $(1,2)$ to $(1, 4.5)$, which does not correspond to any valid byte. The reconstruction step in this case is quite efficient, and it simply corresponds to remapping the modified feature values $(1, 4.5)$ to the closest feature-based representation corresponding to a valid input byte, \ie, $(1, 5)$, which corresponds to byte $1$. 
To summarize, the reconstruction step modifies here the padding bytes (\ie, the red region in Figure~\ref{fig:ramen_scheme}), providing a functional malware sample. However, as this process may not exactly match the optimal feature-based representation, the resulting malware sample may not evade detection or anyway exhibit a lower misclassification confidence.
The feature-space optimization and the reconstruction step can be thus iteratively repeated to improve the attack effectiveness, gradually refining the input malware manipulations. 


\subsection{Gradient-free (Black-box) Attacks}
\label{sect:grad-free}

Another suitable approach to solving Problem~\eqref{eq:ramen} is to use a gradient-free (black-box) optimizer, which only requires querying the target model and observing its outputs, without accessing its internal parameters or knowing how the model works precisely. This approach is also useful when dealing either with non-differentiable models (like decision trees and random forests), for which no gradient information is available, or with feature representations that make the input-space reconstruction step non-trivial or too computationally demanding. This typically happens with hand-crafted features, as those used by the GBDT classifier discussed in Section~\ref{sec:learn-malic-from}, $n$-grams, and other statistical-based features for which there is no clear and direct relationship with the input transformations. 

Within the gradient-free scenario, the attacker must define the \emph{loss function} to be minimized, the \emph{malware manipulations} they will use, and the black-box optimizer that will combine them together.
These steps are similar to the white-box case in Section~\ref{sect:grad-based}, except for the \emph{feature-space optimization} and \emph{input-space reconstruction} steps, which are not required here as the attack is directly optimized in the input space.
Algorithm~\ref{algo:ramen_bb_general} details a high-level implementation of gradient-free attacks.
In each iteration, the attack solves the minimization problem with the chosen optimizer, perturbing the malware with the given practical manipulations (line~\ref{line:bb_only_line}).
In the following, we discuss two different solution strategies, depending on the resources available to the attacker, referred to as \emph{transfer} and \emph{query} attacks.

\begin{algorithm}[t]
    \SetKwInOut{Input}{Input}
    \SetKwInOut{Output}{Output}
    \KwData{$\vct z$, the initial malware sample; $N$, total number of iterations; $y$, the target class label; $f$, the target model function}
    \KwResult{$\vct z^\star$, the adversarial EXEmple.}
    $\vct t^{(0)} \in \faspace$\\
    \textbf{for} $ i $ \textbf{in} $ [0, N-1]$\\
    \Indp
        $\vct t^{(i+1)} = \argmin_{\vct t \in \faspace} L(f(\phi(h(\vct z, \vct t^{(i)}))), y)$\label{line:bb_only_line}\\
    \Indm
    $\vct t^\star = \vct t^{(N)}$\\
    $\vct z^\star = h(\vct z, \vct t^\star)$\\
    return $\vct z^\star$\label{line:ramen_return}
    \caption{Gradient-free (Black-box) Attacks for Optimizing Adversarial Malware EXEmples in \algoname.}
    \label{algo:ramen_bb_general}
\end{algorithm}

\smallskip \noindent \emph{Gradient-free (Black-box) Transfer Attacks.} 
Within this setting, the attacker is assumed to craft the adversarial examples against a \emph{surrogate model}, and then evaluate whether they successfully \textit{transfer} to a different \textit{target model}~\cite{biggio13-ecml,papernot2016transferability,demontis19-usenix,suciu18-usenix,papernot17-asiaccs}.
The surrogate model here is meant to provide a good approximation of the target model, \eg, obtained by training another model on the same classification task. 
The optimization of adversarial EXEmples can be done by following the steps described in Section~\ref{sect:grad-based}, since the surrogate might be differentiable.
The details on how to obtain such a surrogate model are beyond the scope of this paper, but it suffices to say that there are several methods for accomplishing this through model stealing, using open-source models, or simply training a new model on an open-source dataset (\eg, EMBER).
Here we focus on the latter, by optimizing attacks on the networks considered in this work, and transferring them against all the others.
For instance, let us assume that the attacker wants to target a remote service for malware detection (possibly implemented with a DNN-Lin network), and they have access only to the pretrained MalConv classifier. Then, Algorithm~\ref{algo:demetrio_solver} can be used to attack MalConv, while the corresponding adversarial malware samples can be submitted to the target model. Clearly, the success of black-box transfer attacks is inherently connected to the similarity of the decision functions that are learnt by the surrogate and the target models, along with the amount of changes that may be applied to the input sample. We refer the reader to~\cite{demontis19-usenix} for a more comprehensive analysis of the main factors affecting attack transferability between different models.

\smallskip \noindent \emph{Gradient-free (Black-box) Query Attacks.}
Within this scenario, the attacker optimizes directly the malware manipulations by querying the target model and observing its outputs.
To this end, many gradient-free (black-box) optimizers can be exploited, including genetic algorithms~\cite{demetrio2020efficient,castro2019aimed}, natural evolution strategies~\cite{ilyas18-icml,wierstra14-jmlr}, and zeroth-order optimizers~\cite{chen17-aisec}.
Genetic algorithms optimize the objective by maintaining a population of $N$ perturbed samples in each iteration, from which only the best individuals are selected and used to generate the next population. Natural evolution strategies are similar in principle, as they draw the $N$ perturbed samples in each iteration from an underlying Gaussian distribution, and then optimize the parameters of that distribution to minimize the expected loss on the $N$ samples. Zeroth-order optimization aims instead to estimate the gradient of the loss function via finite differences, by querying the target model several times.
All these approaches are iterative, and enable defining a maximum number of queries that can be sent to the target (\ie, a query budget).
As detailed in Section~\ref{subsec:bbox_implement}, we will consider in this work the approach proposed in~\cite{demetrio2020efficient}, which optimizes the padding bytes using a genetic algorithm. 

\subsection{Practical Manipulations}
\label{sec:practical_manipulations}

The optimization algorithms detailed in the previous section can be used to find the best manipulation parameters $\vct t$, which then have to be applied via $h(\cdot, \vct t)$ to craft the actual adversarial malware samples. 
In this section, we enumerate and categorize the \emph{practical manipulations} that can be applied to the input malware and discuss how they can be represented in our framework via the manipulation function $h(\cdot, \vct t)$ and its parameters $\vct t$.

\begin{figure}[t]
    \centering
    \includegraphics[width=0.9\linewidth]{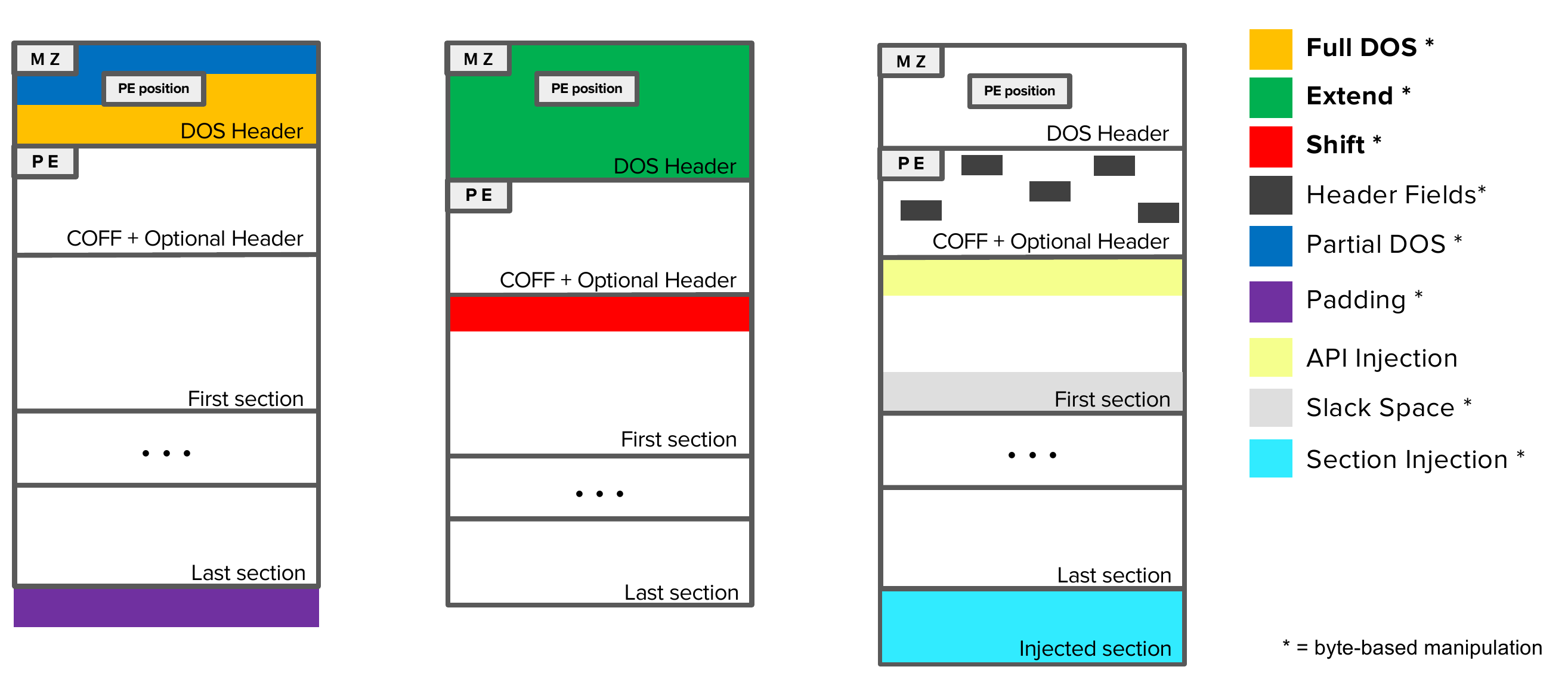}
    \caption{Graphical representation of the locations perturbed by  different attack strategies with the injection of adversarial payloads (shown in colors).
    The three manipulation strategies proposed in this work are highlighted in bold.}
    \label{fig:graphic_manipulations}
\end{figure}

These manipulations are functions that change the representation of a program, by exploiting redundancies and technicalities of the file format, while leaving their functionality intact.
In particular, the attacker aims either to find suitable locations where they can freely alter bytes without breaking the structure, or to create space where they can inject the adversarial payload.
A simplified graphical representation of all these strategies is given in Figure~\ref{fig:graphic_manipulations}.
The colored areas highlight the locations where the adversarial payload can be injected, and the length of the boxes indicates how the content is shifted before injection.

\subsubsection{Novel manipulations}
We discuss here the three novel practical manipulations of Windows programs introduced in this work: \emph{Full DOS}, \emph{Extend}, and \emph{Shift}.
These manipulations are all \textit{byte-based}, \ie, they can manipulate byte values independently. To this end, they exploit ambiguities of the PE format, abusing in-file offsets and adding content to specific locations.
They are described in \algoname by defining $h(\cdot, \vct t)$ as the function manipulating offsets to create space where content can be injected, and $\vct t$ the byte values injected in the predefined locations given by $h(\cdot, \vct t)$.
 
\smallskip \noindent \emph{Full DOS.} This manipulation edits all the bytes contained in the DOS header, which is only kept for compatibility with older operating systems as described in Section~\ref{sec:windows_pe}.
Since the only two important fields in the DOS header are the magic number \verb|MZ| and the 4 byte-long integer at offset \verb|0x3c|, all the other bytes can be used by the attacker to inject the adversarial payload.
Demetrio et al.~\cite{demetrio2019explaining} originally proposed a mutation applied between the magic number and the real header offset, even though, actually, the \textit{whole} DOS header can be manipulated without corrupting the program. In particular, when the program is launched, the control is passed to the loader, which starts parsing the executable. After checking the magic number, it reads the PE offset and it jumps to the metadata of the PE header, skipping the parsing of the DOS header and stub, thus preserving the functionality of the input program.
Algorithm~\ref{algo:full_dos} shows how to perform the editing while keeping the constraints mentioned above.
After that, the algorithm proceeds by rewriting the initial bytes of the DOS header (line~\ref{line:fdos_apply}) and the bytes found after the pointer to the PE header and the PE header itself (line~\ref{line:fdos_pe}).

\begin{algorithm}[t]
    \SetKwInOut{Input}{Input}
    \SetKwInOut{Output}{Output}
    \Input{$\vct z$, the original malware sample; $\vct t$, the vector of parameters}
    \Output{$\vct z^\prime$, the perturbed malware}
    v$ = \vct z.$pe\_offset\\
    $\vct z^\prime = \vct z$\\
    $\vct z^\prime_{2, \ldots, 59} = \vct t_{0, \ldots, 57}$\label{line:fdos_apply}\\
    $\vct z^\prime_{64, \ldots \text{v}-1} = \vct t_{58, \ldots, |\vct t|-1}$\label{line:fdos_pe}\\
    return $\vct z^\prime$
    \caption{Implementation of $h(\vct z, \vct t)$ for the \textit{Full DOS} practical manipulation}
    \label{algo:full_dos}
\end{algorithm}

\smallskip \noindent \emph{Extend.} This manipulation aims to create new space inside the executable by enlarging the DOS header, where the adversary can inject the adversarial noise without breaking the structure of the executable.
To this end, the attacker increases the offset to the PE header, forcing the loader to look up for it further ahead inside the binary, and then they can extend and manipulate this new enlarged DOS header at their will.
Also, the loader will load into memory each header of the program, starting from the DOS header until the end of the Optional header, by looking at a field called \emph{size of headers}.
The latter specifies how many bytes are reserved to the metadata, and it must match their size, rounded up to the nearest multiple of the file alignment.
Hence, to keep intact the structure of the input sample, the attacker must: ($i$) choose an amount that will not collide with the file alignment, ($ii$) enlarge the field containing the size of the headers, and ($iii$) increment the offset of each section entry.
After this editing, the loader will again skip the adversarial content in search for the PE header, without altering its execution flow.
Algorithm~\ref{algo:extend} shows the aforementioned approach.
It first computes the amount of bytes to be added from the vector of parameters (line~\ref{line:inj}), and then it retrieves the PE header offset (line~\ref{algo:pe_offset}) to create the space that will host the adversarial payload (line~\ref{line:ext_fill}).
With a little abuse of notation, we write \texttt{0x00}*\emph{inj}, implying that the single byte \texttt{0x00} is concatenated with itself \emph{inj} times, and we use the symbol \texttt{+} to denote the concatenation of byte strings and byte values.
Then, the algorithm fixes all the constraints imposed by the format: (\textit{i}) it increases the offset to the PE header (line~\ref{line:pe_adj}), (\textit{ii}) it increases the \emph{size of headers} field (line~\ref{line:pe_szheaders}), and  (\textit{iii}) it increases all the section entries (line~\ref{line:section_entries_ext}).
Lastly, the algorithm perturbs all the allowed bytes inside the enlarged DOS header (line~\ref{line:extend_dos_copy} and~\ref{line:extend_t_copy}), similarly to Algorithm~\ref{algo:full_dos}.

\begin{algorithm}[t]
    \SetKwInOut{Input}{Input}
    \SetKwInOut{Output}{Output}
    \Input{$\vct z$, the original malware sample; $\vct t$, the vector of parameters}
    \Output{$\vct z^\prime$, the perturbed malware}
    \emph{inj} $=|t| - 58$\label{line:inj}\\
    v$ = \vct z.$pe\_offset\label{algo:pe_offset}\\
    $\vct z^\prime = \vct z_{0, \ldots, \text{v}-1} + $ \texttt{0x00} * \emph{inj} $ + \vct z_{\text{v}, \ldots, |\vct z|-1}$\label{line:ext_fill}\\
    $\vct z^\prime$.pe\_offset = v + \emph{inj}\label{line:pe_adj}\\
    $\vct z^\prime$.sizeof\_headers = $\vct z$.sizeof\_headers + \emph{inj}\label{line:pe_szheaders}\\
    \emph{S} $= \vct z^\prime.$get\_sections()\\
    \For{\emph{s} \textbf{in} \emph{S}}{
        $s.$physical\_offset =  \emph{s}.physical\_offset + \emph{inj}\label{line:section_entries_ext}\\
    }
    $\vct z^\prime_{2, \ldots, 59} = \vct t_{0, \ldots, 57}$\label{line:extend_dos_copy}\\
    $\vct z^\prime_{64, \ldots \text{v}-1} = \vct t_{58, \ldots, |\vct t|-1}$\label{line:extend_t_copy}\\
    return $\vct z^\prime$
    \caption{Implementation of $h(\vct z, \vct t)$ for the \textit{Extend} practical manipulation}
    \label{algo:extend}
\end{algorithm}

\smallskip \noindent \emph{Shift.} This manipulation aims to create new space inside the executable by shifting the content of the first section and injecting there the adversarial noise.
We recall from Section~\ref{sec:windows_pe} that each section entry specifies an offset inside the binary where the loader can find the content of that section.
Each offset is multiple of the file alignment, specified in the Optional Header.
The job of the loader is to read these entries, and load into memory the content of each section, by looking inside the binary at the offset specified by the entry.
The attacker can not interfere with this behaviour, but they can increase the offset of each section entry to carve space before the beginning of each section, and fill the new crafted holes with chunks of bytes.
As discussed before, the injected content must be, in size, multiple of the file alignment, to keep the structure of the program intact.
In this way, the loader will look for the content of each section, ignoring the adversarial perturbation introduced between them.
For the scope of this paper, we only inject content before the first section, but in principle this technique could also be used for shuffling the order of the sections inside the binary, or also inject different-in-length chunks of bytes between sections.
Algorithm~\ref{algo:shift} shows this approach.
It retrieves the position of the first section inside the binary (line~\ref{line:fs_offset}), and it inserts the desired number of bytes in that position (line~\ref{line:sht_fill}).
Then, the algorithm fixes each offset of the section entries of the program (line~\ref{line:section_entries_ext}), and it copies the adversarial payload in the newly-created space (line~\ref{line:shift_t_copy}).

\begin{algorithm}[t]
    \SetKwInOut{Input}{Input}
    \SetKwInOut{Output}{Output}
    \Input{$\vct z$, the original malware sample; $\vct t$, the vector of parameters}
    \Output{$\vct z^\prime$, the perturbed malware}
    \emph{inj} $=|t|$\label{line:inj}\\
    $\nu = \vct z.$get\_first\_section\_offset()\label{line:fs_offset}\\
    $\vct z^\prime = \vct z_{0, \ldots, \text{v}-1} + $ \texttt{0x00} * \emph{inj} $ + \vct z_{\text{v}, \ldots, |\vct z|-1}$\label{line:sht_fill}\\
    $S =$ $\vct z^\prime.$get\_sections()\label{line:for_b}\\
    \For{$s $ \textbf{in} $ S$}{
        $s.$physical\_offset + \emph{s}.physical\_offset + \emph{inj}\\
    }\label{line:end_f}
    $\vct z^\prime_{ {\text{v}}, \ldots, \text{v+(inj-1)}} = \vct t$\label{line:shift_t_copy}\\
    return $\vct z^\prime$
    \caption{Implementation of $h(\vct z, \vct t)$ for the \textit{Shift} practical manipulation}
    \label{algo:shift}
\end{algorithm}

\subsubsection{Previously-proposed Practical Manipulations.} We provide here an overview of the previously-proposed practical manipulations applicable to PE files which are encompassed by our framework~\cite{anderson2017evading, castro2019aimed, kolosnjaji2018adversarial, demetrio2019explaining, demetrio2020efficient, song2020automatic, sharif2019optimization, kreuk2018deceiving, suciu2019exploring, demetrio2020efficient}. 

\myparagraph{Byte-based Manipulations.} We start by discussing \textit{byte-based manipulations} that can alter byte values independently.

\smallskip \noindent \emph{Header Fields}~\cite{anderson2017evading}.
This manipulation technique modifies specific fields contained inside the PE and the Optional Header, \eg, changing the name of each section by editing the corresponding section entries. Each byte in these fields can be independently and arbitrarily changed.

\smallskip \noindent \emph{Partial DOS}~\cite{demetrio2019explaining}.
This manipulation alters the first 58 bytes of the unused DOS header, starting from the byte after the magic number \texttt{MZ} to the offset to PE header. 

\smallskip \noindent \emph{Slack Space}~\cite{kreuk2018deceiving,suciu2019exploring}.
This manipulation fills the space between sections. 
When the program is compiled, to keep the beginning of each section aligned to the value specified in the header, the compiler appends a trail of zero bytes to each section to fill the gap. This space can be used to inject the adversarial payload.

\smallskip \noindent \emph{Padding}~\cite{kolosnjaji2018adversarial,kreuk2018deceiving}.
This manipulation appends padding bytes at the end of the input file.

\smallskip \noindent \emph{Section Injection}~\cite{anderson2017evading, demetrio2020efficient}.
This manipulation creates a new section that is injected inside the target executable, along with a new section entry inside the Section Table. Accordingly, this manipulation does not only change the byte distribution exhibited by the input program, but also its structure.

\myparagraph{Other Manipulations.} We finally discuss other practical manipulations which cannot manipulate byte values independently (\ie, they are not byte-based), as this would corrupt the input file functionality.

\smallskip \noindent \emph{API Injection}~\cite{anderson2017evading}.
This manipulation aims to add entries inside the Import Table of an executable, causing the OS to include more APIs during the loading process.
While the attacker can not remove APIs, as this would break the functionality of the program, they can inject API imports that will not be used by the program.
In this case, the injection is made up of a complete entry that can not contain arbitrary values, but rather it must comply with a specific format.

\smallskip \noindent \emph{Binary Rewriting}~\cite{wenzl2019hack, sharif2019optimization}.
These manipulations allow the attacker to alter the code of the program by different means, spanning from substituting a set of instructions with others that are semantically equivalent (\eg replacing additions with subtraction, and changing signs of the values), to encode all the program inside another one (packing). It is thus clear that, also in this case, the input bytes corresponding to the location affected by this manipulation can not be changed independently from each other.
We refer the reader to~\cite{demetrio2020efficient} for more details on how binary-rewriting techniques can be used to craft adversarial malware.

\subsection{Implementing Practical Gradient-based (White-box) Attacks within \algoname}
\label{subsec:wbox_implement}

In this section, we discuss how to implement gradient-based (white-box) attacks within \algoname, by connecting the implementation of Algorithm~\ref{algo:ramen_wb_general} (and Figure~\ref{fig:ramen_scheme}) with the corresponding practical manipulations (described in Section~\ref{sec:practical_manipulations}) used by each attack. 
To this end, for each attack, we have to define the following components: (\textit{i}) the loss function, (\textit{ii}) the considered practical manipulations, (\textit{iii}) how the gradient is exploited to perform the feature-space optimization, and (\textit{iv}) how the adversarial malware is eventually reconstructed in the input space.
Table~\ref{tab:ramen_related} summarizes such specifications both for our novel attacks and for previously-proposed ones, as detailed more specifically in the following.
\begin{table}[t]
\resizebox{\textwidth}{!}{%
\begin{tabular}{@{}lllll@{}}
\toprule
    Attack &
    Loss Function $L$ &
    Practical Manipulations $h(\cdot, \vct t)$ &
    Feature-space Optimization &
    Input-space Reconstruction \\
    \midrule
Full DOS (this work) &
    malware score &
    manipulate all DOS header &
    single gradient step &
    closest positive (iterative) \\
Extend (this work) &
    malware score &
    extend DOS header &
    single gradient step &
    closest positive (iterative) \\
Shift (this work) &
    malware score &
    shift section content &
    single gradient step &
    closest positive (iterative) \\
\hline
Padding~\cite{kolosnjaji2018adversarial} &
    malware score &
    padding &
    single gradient step &
    closest positive (iterative) \\
Partial DOS~\cite{demetrio2019explaining}&
    malware score &
    partial DOS header &
    single gradient step &
    closest positive (iterative) \\
FGSM~\cite{kreuk2018deceiving, suciu2019exploring}&
    malware score &
    padding + slack space &
    FGSM &
    closest (non-iterative) \\

Binary Diversification~\cite{sharif2019optimization}&
    CW loss & 
    equivalent instructions &
    single gradient step &
    gradient-aligned transformation (iterative) \\
  \bottomrule
\end{tabular}}
\caption{Implementing novel and state-of-the-art gradient-based (white-box) attacks within \algoname, according to the steps detailed in Algorithm~\ref{algo:ramen_wb_general} and Figure~\ref{fig:ramen_scheme}.}
\label{tab:ramen_related}
\end{table}

\subsubsection{Byte-based Attack Algorithm}  For attacks using byte-based manipulations against DNNs trained on byte embeddings (such as MalConv and the DNNs discussed in Section~\ref{sec:learn-malic-from}), Algorithm~\ref{algo:ramen_wb_general} can be concretely implemented as detailed in Algorithm~\ref{algo:demetrio_solver}.
The main idea is to iteratively update the byte values $\vct t$ that can be changed by the given byte-based manipulation (\eg, the padding bytes) according to the gradient computed in the embedding space.

The attack (Algorithm~\ref{algo:demetrio_solver}) works as follows. As in~\cite{kolosnjaji2018adversarial}, our attack starts by creating a temporary matrix containing all the embedding values, using $\hat{\phi}$ that is the function that encodes a single byte inside the feature space (line~\ref{line:embedding}).
It initializes the initial vector of parameters $\vct t$ with random values (line~\ref{line:t_init}), and it starts the optimization.
At the beginning of each iteration, the algorithm applies the practical manipulations optimized so far and it encodes the sample inside the feature space (line~\ref{line:apply_phi}).
Then, the optimizer computes the gradient \wrt the embedded bytes (line~\ref{line:gradient}), which is negative here as we are considering a minimization problem.
Note also that the gradient is a matrix in this case, given that each byte is represented as a vector in the embedding space.
The algorithm ignores all the locations that cannot be modified by applying a binary mask $\vct m$ to the gradient (line~\ref{line:gradient}).
In this way, the algorithm will discard every location that is not related to the computation of the attack, \ie, bytes which are not associated to the perturbed input bytes $\vct t$.
After applying the mask, the algorithm computes the norm of each embedded value (line~\ref{line:norms}) and it takes the indexes of the first $\gamma$ non-zero sorted entries (line~\ref{line:entries}).
The $\gamma$ parameter is a step-size constant that controls how many bytes are perturbed at each round, modulating how much the space is explored during the optimization.
For each perturbed byte, the algorithm computes a line passing from the current value to be replaced, and whose direction is imposed by the gradient in that point.
The algorithm proceeds by projecting all the 256 embedded byte values on this direction, computing the distance point-to-line and the alignment with such direction (line~\ref{line:distance}).
Hence, our optimizer always performs one \emph{single gradient step} at the time.
Finally, the algorithm iterates over all the computed distances, and it chooses the byte associated with the closest embedding value that lies on the positive direction pointed by the gradient (line~\ref{line:argmin}).
Since each embedding value computed in this way has a one-to-one mapping with the bytes in input space, and only values aligned with the gradient will be chosen, the reconstruction is \emph{closest positive}, since they must also be on the positive side of the privileged direction.
Our approach is \emph{iterative}, after $N$ iterations the hybrid optimizer returns the adversarial EXEmples $\vct z^\star$ optimized so far.

\begin{algorithm}[t]
    \SetKwInOut{Input}{Input}
    \SetKwInOut{Output}{Output}
    \Input{$\vct z$ the original malware sample; $\gamma$ the number of bytes to optimise; $N$ the total number of iterations}
    \Output{$\vct z^\star$, the adversarial EXEmple}
    $\mat E_i = \hat \phi(i), \forall i \in [0, 256]$\label{line:embedding}\\
    $\vct t^{(0)} \in \faspace$\label{line:t_init}\\
    \textbf{for} $ i $ \textbf{in} $ [0, N-1]$\\
    \Indp
        $\mat X \leftarrow \phi(h(\vct z, \vct t^{(i)}))$ // feat. space \label{line:apply_phi}\\
        $\mat G \leftarrow -\nabla_{\mat X} f (\mat X) \odot \vct m$\label{line:gradient}\\
        $\vct g \leftarrow (\parallel \mat G_0 \parallel, ... ,\parallel \mat G_n \parallel)$\label{line:norms}\\
        \textbf{for} $ k $ \textbf{in} $ \text{argsort~}(\vct g)_{0, ..., \gamma} \wedge g_k \neq 0$\label{line:entries}\\
        \Indp 
            \textbf{for} $ j $ \textbf{in} $ [0, ..., 255]$\\
            \Indp
                $\mat S_{k, j} \leftarrow \mat G_k^t \cdot (\mat E_j - \mat X_k)\quad$\\
                $\widetilde{\mat X}_{k, j} \leftarrow \parallel \mat E_j - (\mat X_k + \mat G_k \mat S_{k, j})\parallel_2 $\label{line:distance}\\
            \Indm
        $\vct t^{(i+1)}_k \leftarrow  \argmin_{j : \mat S_{k, j} > 0} \widetilde{\mat X}_{k, j}$ \label{line:argmin} //input space\\
        \Indm
    \Indm
    $\vct t^\star \leftarrow \vct t^{(N)}$\\
    $\vct z^\star \leftarrow h(\vct z, \vct t^\star)$\\
    \text{\textbf{return}} $\vct z^\star$
    \caption{Implementation of Algorithm~\ref{algo:ramen_wb_general} for byte-based attacks on DNNs using byte embedding}
    \label{algo:demetrio_solver}
\end{algorithm}

\subsubsection{Novel attacks}
We introduce here the three attacks proposed in this work, \ie, \textit{Full DOS}, \textit{Extend}, and \textit{Shift}. 
All of them rely on Algorithm~\ref{algo:demetrio_solver} for manipulating each single byte that is injected and manipulated, by specifying the practical manipulation they apply and how they shape the mask $\vct m$. 

\myparagraph{Full DOS.}
To implement this attack, we use the homonym manipulation as the $h$ function (line~\ref{line:apply_phi}) described in Algorithm~\ref{algo:full_dos}, and we customize the mask $\vct m$ by specifying as editable (\textit{i}) the first 58 bytes after the magic number \texttt{MZ}, and (\textit{ii}) all the bytes from the pointer to the PE header, to the beginning of the PE header itself.

\myparagraph{Extend.}
To implement this attack, we use the homonym manipulation as the $h$ function (line~\ref{line:apply_phi}) described in Algorithm~\ref{algo:extend}, that enlarge the DOS header as specified by the vector of manipulation $\vct t$.
Hence, the mask $\vct m$ includes all the locations manipulated by the \emph{Full DOS} attack, plus all the content that has been injected by the practical manipulation.

\myparagraph{Shift.}
To implement this attack, we use the homonym manipulation as the $h$ function (line~\ref{line:apply_phi}) described in Algorithm~\ref{algo:shift}, that shifts the content after the PE header as specified by the vector of manipulation $\vct t$.
Hence, the mask $\vct m$ specify as editable all the locations that are added by the practical manipulation.

\subsubsection{Recasting Existing Attacks into \algoname}
\label{sec:soa_implementation}
After having shown how \algoname can encode many different types of practical manipulations, we reinforce the claim of its generality by showing how to encode also other optimization strategies proposed in the \SoA.

\myparagraph{Padding.} Kolosnjaji et al.~\cite{kolosnjaji2018adversarial} optimize the generation of adversarial EXEmples by minimizing the malicious score computed by the classifier, and they apply the \emph{Padding} practical manipulation for injecting the content inside the input samples.
The optimizer they use is is gradient-based, and it is similar to Algorithm~\ref{algo:demetrio_solver}, as they choose the closest embedding value in the positive direction of the gradient (\emph{closest positive} in Table~\ref{tab:ramen_related}).
Lastly, the reconstruction inside the input space is done by inverting the embedding look-up that is applied on bytes (\emph{inverse look-up} in Table~\ref{tab:ramen_related}).
This is doable, since the optimizer only chooses embedding values that correspond to real bytes, and the one-to-one mapping is preserved.

\myparagraph{Partial DOS.} Similarly, also Demetrio et al.~\cite{demetrio2019explaining} use the malicious as loss function to minimize. 
The manipulation they use is the \emph{Partial DOS}, and again they use the same optimizer and reconstruction algorithm as Kolosnjaji et al.~\cite{kolosnjaji2018adversarial}.

\myparagraph{FGSM.} Kreuk et al.~\cite{kreuk2018deceiving} and Suciu et al.~\cite{suciu2019exploring} use the malware score as loss function to minimize, and both of them rely on the \emph{Slack Space} and \emph{Padding} manipulations.
They leverage a gradient technique inspired to the Fast Gradient Sign Method (FGSM)~\cite{goodfellow2014explaining}: they first map the sample inside the feature space, and they apply the classical FGSM method, moving the feature vectors without applying constraints.
The latter is formalized as $\vct x_{i+1} = \vct x_{i} + \epsilon sign(\nabla_{\vct x_{i}} f(\vct x_{i}))$, where $\epsilon$ is the maximum distortion that can be applied on the sample.
While Suciu et al.~\cite{suciu2019exploring} use only one single iteration for this technique, Kreuk et al.~\cite{kreuk2018deceiving} performs more steps inside the feature space, repeating the process until evasion.
Since the non-iterative version is just the first step of the iterative one, we will discuss only the latter, since it is more generic.
The reconstruction is done at the end of the optimization, where each embedding value is must be converted to a real byte inside the input space (\emph{closest} in Table~\ref{tab:ramen_related}).
This is done for each entry of the final feature vector, by taking the byte whose embedding value is closest to the one contained inside the embedded malware.

\myparagraph{Binary Diversification.} Sharif et al.~\cite{sharif2019optimization} use the Carlini \& Wagner loss (CW loss)~\cite{carlini2017towards}, and they
apply random manipulations that change instructions inside the \emph{.text} section with semantics-equivalent ones, or they displace the code inside another section, with the use of \emph{jump} instructions.
At each iteration of the algorithm, the latest adversarial example is used as a starting point for the new one.
Once randomly perturbed, the new and the old versions are projected inside the embedding space, where the two points are used for computing a direction.
If this direction is parallel to the gradient of the function in that point, the sample is kept for the next iteration, otherwise it is discarded.
In this case, the strategy does not optimize the sample inside the feature space, since each sample is constructed by applying random transformation.

Except for the manipulations proposed by Sharif et al.~\cite{sharif2019optimization}, all the strategies described so far can be found in the \texttt{secml-malware} library released along with this paper.\footnote{\url{https://github.com/zangobot/secml_malware}}

\subsection{Implementing Practical Gradient-free (Black-box) Attacks within \algoname}
\label{subsec:bbox_implement}

We discuss here how to implement gradient-free (black-box) attacks within \algoname based on detailing the implementation of Algorithm~\ref{algo:ramen_bb_general} using the practical manipulations described in Section~\ref{sec:practical_manipulations}.

To this end, one has to define the following components: (\textit{i}) which loss function they minimize, (\textit{ii}) which practical manipulations they use, and (\textit{iii}) which optimizer they apply for solving the problem.
We fill Table~\ref{tab:bb_ramen_related} with all the specifications used for implementing either our novel attacks, either the existing techniques proposed in the previous literature within the \algoname framework.
\begin{table}[t]
\resizebox{\textwidth}{!}{%
\begin{tabular}{@{}llllll@{}}
\toprule
    Attack &
    Loss Function $L$ &
    Practical Manipulations $h(\cdot, \vct t)$&
    Optimizer &
    Validation \\
    \midrule
Full DOS (this work) &
    malware score &
    manipulate all DOS header &
    genetic &
    none \\
Extend (this work) &
    malware score &
    extend DOS header &
    genetic &
    none \\
Shift (this work) &
    malware score &
    shift section content &
    genetic &
    none \\
Partial DOS (this work) &
    malware score &
    partial DOS header &
    genetic &
    none \\
Padding (this work) &
    malware score &
    padding &
    genetic &
    none \\
    \midrule
GAMMA padding~\cite{demetrio2020efficient} &
    malware score + size penalty &
    padding with benign sections / benign section injection&
    genetic &
    none \\
RL Agent~\cite{anderson2017evading} &
    malware score&
    padding + section / API inj. + header fields + binary rewriting&
    reinforcement learning &
    none \\
AIMED~\cite{castro2019aimed} &
    malware score&
    padding + section/API inj. + header fields + binary rewriting&
    genetic &
    sandbox \\
AEG~\cite{song2020automatic} &
    malware score&
    padding + section inj. + header fields + binary rewriting&
    random manipulations &
    sandbox \\
  \bottomrule
\end{tabular}}
\caption{Implementing novel and state-of-the-art gradient-free (black-box) attacks within \algoname, according to the steps detailed in Algorithm~\ref{algo:ramen_bb_general}.}
\label{tab:bb_ramen_related}
\end{table}

\subsubsection{Gradient-free (black-box) byte-based optimizer}

Symmetrically to Section~\ref{subsec:wbox_implement}, we start by introducing the optimizer we developed for this work.
Since no gradient information is available, this optimizer relies on a genetic algorithm to compute bytes that mostly decreases the chosen loss function, injected by applying practical manipulations.
We extend \algoname with the genetic black-box optimizer used by Demetrio et al.~\cite{demetrio2020efficient} for computing their attack.
This algorithm creates byte sequences that are injected inside the sample through the usage of practical manipulations, and the newly generated adversarial EXEmple is sent to the detector to be scored.
Also, this optimizer is bounded by the total number of queries that can be sent to the detector, and the process halts when the algorithms has reached that limit.
Since this genetic black-box optimizer works with real numbers, but the values that we inject are integer values between \texttt{0} and \texttt{255}, we encode our vector of parameters as $\vct t \in [0, 1]^k$, where $k$ is the number of values that will be perturbed (\eg the \emph{Partial DOS} attack sets $k$ to \verb|58|), and we multiply this value by \texttt{255} when the vector of parameter is passed to the $h$ function.
Before applying the practical manipulation $h$, we need to multiply by \verb|255| and rounding to the nearest value the vector of parameter, since Algorithm~\ref{algo:full_dos},~\ref{algo:extend} and~\ref{algo:shift} consider each entry of vectors $\vct t$ as bytes to be placed inside the sample.
We summarize the optimizer in Algorithm~\ref{algo:genetic_algorithm}, where we have plugged the loss to minimize as dictated by \algoname.
The algorithm is initialized by randomly generating a matrix of bytes sequences to inject $\mat S^\prime = (\vct t_{1}, ..., \vct t_{N}) \in \faspace^N \subset [0,1]^{N \times k}$, which represents the initial population of $N$ candidate manipulation vectors (line~\ref{line:initial_population}). 
The genetic optimizer iterates over three steps that ensures the differentiation of the upcoming generation of candidates: \emph{selection}, \emph{cross-over}, and \emph{mutation}.
The \emph{selection} step (line~\ref{line:selection}) applies the objective function $F$ to evaluate the candidates in $\mat S^\prime$.
It then selects the best $N$ candidates between the current population $\mat S^\prime$ and the population generated at the previous iteration $\mat S$.
The resulting vectors are the candidates associated with the lowest values of $F$, \ie the most promising for creating adversarial EXEmples.
The \emph{crossover} function (line~\ref{line:crossover}) takes the selected candidates as input and returns a novel set of $N$ candidates by mixing the values of pairs of randomly-chosen vector candidates.
The \emph{mutation} function (line~\ref{line:mutation}) changes the elements of each input vector at random, with a fixed low probability.
The combination of both \emph{cross-over} and \emph{mutation} ensures that each population has different traits \wrt to the previous one, allowing the algorithm to properly chose new potential solution, \ie byte sequences, discarding the useless ones, and exploring the space of feasible solutions.
After $\vct T$ queries, the optimizer extracts the vector of manipulation $\vct t$ associated with a minimal value for the objective function $F$ (line~\ref{line:best_t}), and it is used to create and return the final adversarial EXEmple $\vct z^\star$ (line~\ref{line:genetic_end}).

\begin{algorithm}[t]
    \SetKwInOut{Input}{Input}
    \SetKwInOut{Output}{Output}
    \Input{$\vct z$, the initial malware sample; $N$, the population size; $T$, the query budget.}
    \Output{$\vct z^\star$, the adversarial EXEmple.}
    $q \leftarrow 0$, $\mat S \leftarrow \emptyset$ \\
    $\mat S^\prime \leftarrow (\vct t_1, ..., \vct t_N) \in \faspace^N$ \label{line:initial_population}\\
    \While{ $ q < T$}{
    \label{algo:loop}
        $\mat S \leftarrow \text{selection}(\mat S \cup \mat S^\prime, F, \vct t)$ \label{line:selection}\\
        $\mat S^\prime \leftarrow$ crossover($\mat S$)\label{line:crossover}\\
        $\mat S ^\prime\leftarrow$  mutate($\mat S^\prime$) 
        \label{line:mutation}\\
        $q \leftarrow q + N$\\
    }
    $\vct t^\star \leftarrow$, best candidate from $\mat S$ with minimum $F$\label{line:best_t}\\
    $\vct z^\star \leftarrow h(\vct z, \vct t^\star)$\\
    \textbf{return} $\vct z^\star\label{line:genetic_end}$
    \caption{Implementation of Algorithm~\ref{algo:ramen_wb_general} for implementing byte-based attacks against Windows malware classifiers}
    \label{algo:genetic_algorithm}
\end{algorithm}

\subsubsection{Novel attacks}
Since the algorithm optimizes directly $\vct t$, the attacker only need to chose a practical manipulation of their choice, and plug it inside the objective function $F$.
In this work, the attacks that we implement with Algorithm~\ref{algo:genetic_algorithm} are \emph{Full DOS}, \emph{Extend}, \emph{Shift}, \emph{Partial DOS}, and \emph{Padding}, and all of them use the malware score as loss function to minimize.

\subsubsection{Recasting Existing Attacks into \algoname}
Here we show how can we re-implement existing attacks into \algoname, by specifying the main components of out formalization for each of them.

\myparagraph{GAMMA}~\cite{demetrio2020efficient}.
Formulated as a black-box query attack, it tries to find adversarial EXEmples by minimizing the malware score, plus a penalty term that controls the size of the injected content: $f(\phi(h(\vct z, \vct t)) + \lambda C(\vct t) $, where $\lambda$ is the regularization parameter, and $C$ is a function that counts how many bytes have been injected using $\vct t$.
In this scenario, each vector $\vct t$ specifies how much content must be harvested from goodware programs, injected using either the \emph{Padding} or the \emph{Section Injection} practical manipulation.
Since these manipulations do not break the functionality by design, the attack does not need a sandbox that verifies the functionality of the perturbed malware.
Then, the attack applies a genetic optimizer to find a minimum to the aforementioned function, by taking into account both the score and the size constraint.

\myparagraph{RL Agent}~\cite{anderson2017evading}.
Formulated as a black-box query attack, this strategy minimizes the malware score by using a reinforcement learning algorithm.
It uses a mixture of many practical manipulations (\emph{Padding}, \emph{Section Injection}, \emph{API Injection}, \emph{Header Fields}) with random content, and the agent is trained is trained to evade a local model (\ie a baseline version of the GBDT classifier).
They also use one single \emph{Binary rewriting} manipulation, that creates a new entry point inside the code of the program. 
The new code then jumps back to the original entry point, continuing the normal execution.
It learns the best sequence of actions that leads to adversarial EXEmples, and then it is tested against other targets (\eg VirusTotal).
They do not use any sandbox, since most of the manipulations are functionality-preserving except for the entry point modification.
The latter could break the functionality, since the program could be loaded at a custom address in memory, hence the jump could fail to find the original code, causing a crash.

\myparagraph{AIMED}~\cite{castro2019aimed}.
This strategy is very similar to the one applied by the \emph{RL agent}~\cite{anderson2017evading} discussed earlier.
They minimize the malware score, and they use the exact same manipulations in the same way.
The difference is in the way they optimize the chain of manipulations: here, they use a genetic optimizer that looks for the best sequence of actions to apply.
At each step of the optimization process, they validate the newly created adversarial EXEmple by testing it inside a sandbox.

\myparagraph{Adversarial Example Generation (AEG)}~\cite{song2020automatic}.
This strategy minimizes the score of the target classifier, and it applies a mixture of different practical manipulations (\emph{Padding}, \emph{Section Injection}, \emph{Header Fields}), and they also apply one \emph{Binary rewriting} manipulation, that is the swapping of assembly instructions with equivalent ones (similarly to Sharif et al.~\cite{sharif2019optimization}).
Then, they apply these manipulations at random, looking for the best sequence of actions that leads to evasion.
At each step, the attack checks the validity of the current sample inside a sandbox environment.

\section{Experimental Analysis}
\label{sec:experiments}
\label{sec:setup}

\begin{figure}[t]
    \centering
    \includegraphics[width=0.95\linewidth]{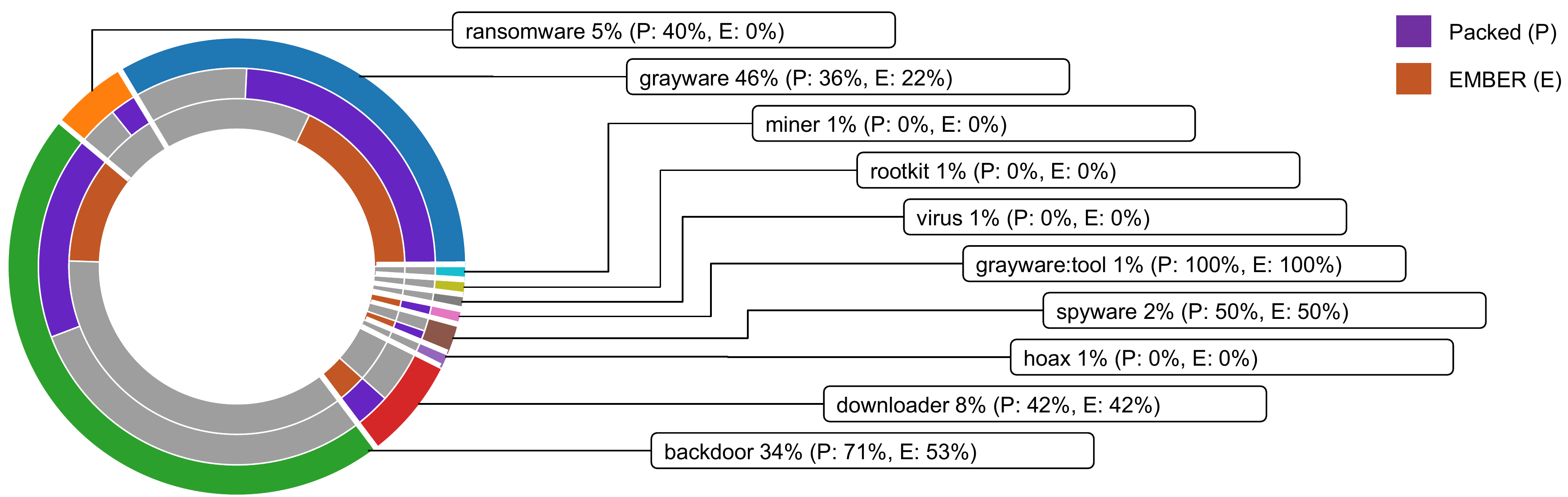}
    \caption{The composition of the test dataset used during the attacks. The legend shows also the percentage of packed samples (\textbf{P}) in each malware category, and the percentage of malware present in the EMBER dataset (\textbf{E}).}
    \label{fig:malware_pie}
\end{figure}

We used a Ubuntu 16.04.3 LTS server, with an Intel\textsuperscript{\textregistered} Xeon\textsuperscript{\textregistered} E5-2630 CPU, and 64 GB of RAM to perform our experiments.
We also used a Windows 10 virtual machine during the development of the practical manipulations described in Section~\ref{sec:practical_manipulations} to debug them during their development, and to validate that they indeed do not impact functionality.
This virtual machine has not been used to test all the samples, but only for studying the manipulations and the file format with the current version of the Windows loader.
To highlight the performance of our strategy, we encoded other attacks proposed in the \SoA ~\cite{demetrio2019explaining, kolosnjaji2018adversarial, suciu2019exploring, kreuk2018deceiving} and we ran them against the chosen targets.
The network proposed by Johns\footnote{\url{https://www.camlis.org/2017/jeffreyjohns}} and Coull et al.~\cite{coull2019activation} has been trained with two different datasets.
The first one is EMBER~\cite{anderson2018ember}, which is an open-source dataset of goodware and malware hashes, including a set of pre-extracted features, while the second is a proprietary production-quality dataset used for training malware classifiers.
The first dataset is smaller, counting \embersize samples, while the second is larger, counting \largedatasetsize files.
MalConv has been trained on EMBER~\cite{anderson2018ember}, like the GBDT model proposed by Anderson et al.~\cite{anderson2018ember}.
We encoded all the strategies inside a Python library we are developing and maintaining, called \texttt{secml-malware},~\footnote{\url{https://github.com/zangobot/secml_malware}} as an extension of \texttt{secml}~\cite{melis2019secml}.
Regarding the black-box setting, we omit from the analysis the FGSM proposed by Kreuk et al.~\cite{kreuk2018deceiving} and Suciu et al.~\cite{suciu2019exploring}, since they are similar to the Padding strategy.
We also test GAMMA~\cite{demetrio2020efficient}, a regularized black-box strategy whose practical manipulations consist in injecting content harvested from benign software to elude detection.
For this experimental setup, we rely on the \emph{padding} version of this attack, where the adversarial noise is appended at the end of the input malware.

\myparagraph{Test Dataset.}
The malware set we used for the empirical evaluation is a richer version of the one used by Demetrio et al.~\cite{demetrio2019explaining}, and we show its composition in Figure~\ref{fig:malware_pie}.
Each slice of the plot resembles the percentage of one or more families of malware. 
The labels of the legend also depict the percentage of samples of that families that are packed, and how many of them are also contained inside the EMBER dataset.
These samples were retrieved from \emph{DasMalwerk}~\footnote{\url{https://www.dasmalwerk.eu/}} during late 2018.
The total amount of samples is \testdataset, and \inember of them are contained inside the EMBER dataset (roughly, one third of the overall dataset).
This dataset has been labelled by querying VirusTotal,~\footnote{\url{https://www.virustotal.com}} and it includes 46\% of \textit{grayware}, 33\% of \textit{backdoor}, 5\% of \textit{ransomware}, and smaller fractions of \textit{spyware}, \textit{rootkits} and \textit{miners}.
According to the analysis, \packed of them are packed.

\begin{figure}[t]
\begin{subfigure}{.45\textwidth}
    \centering
    \includegraphics[width=\textwidth]{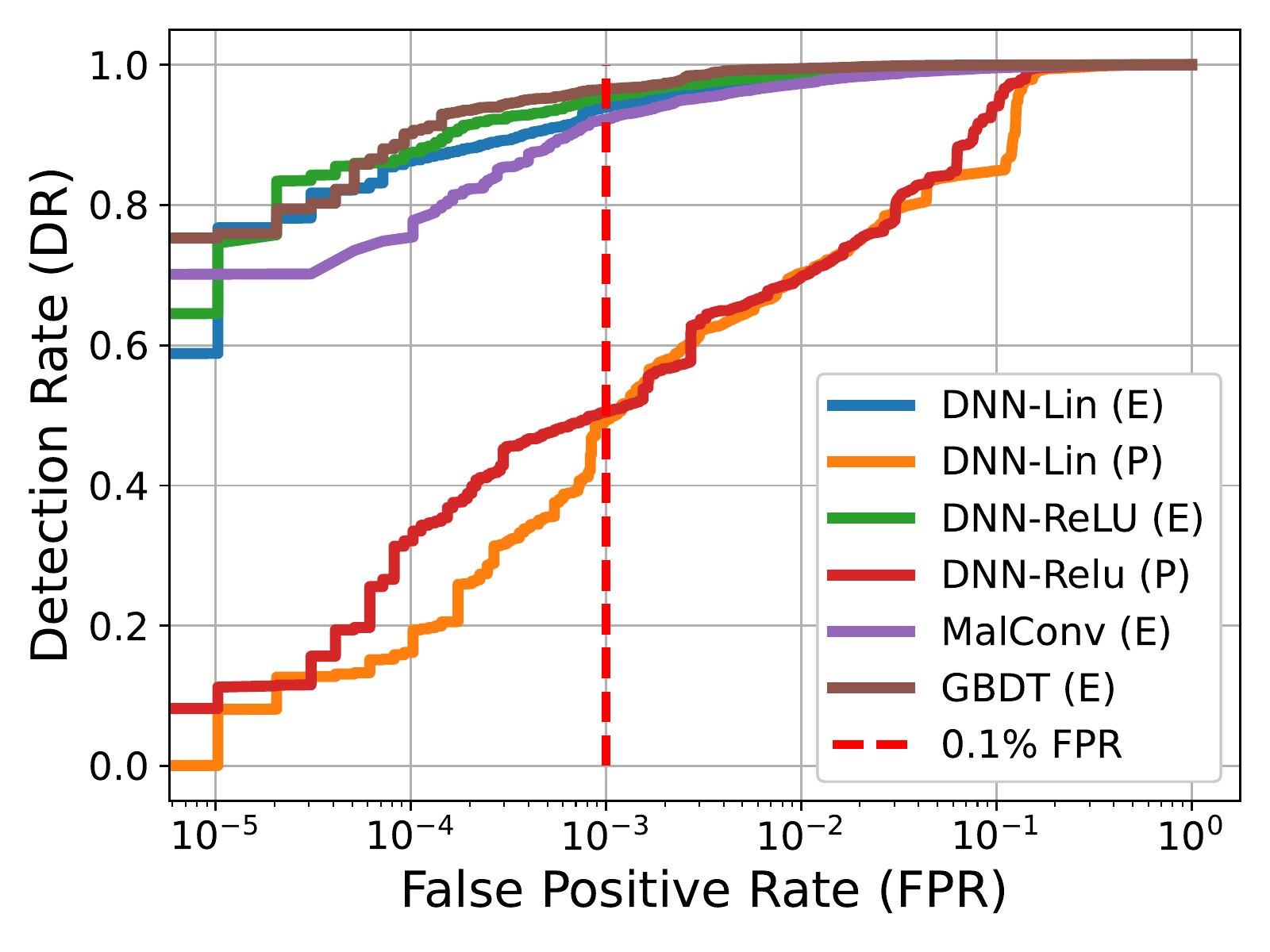}
    \caption{}
    \label{fig:roc}
\end{subfigure}
\begin{subfigure}{.4\textwidth}
\centering
    \begin{tabular}{c|c|c}
        \textbf{Model} & \textbf{$\vct \theta$ (at 0.1\% FPR)} & \textbf{DR (at 0.1\% FPR)}\\\hline
        DNN-Lin (E)& 0.97 & 0.93 \\\hline
        DNN-Lin (P)& 0.94 & 0.49 \\\hline
        DNN-ReLU (E) & 0.99 & 0.95\\\hline
        DNN-ReLU (P) & 0.92 & 0.5 \\\hline
        MalConv (E) & 0.99 & 0.92 \\\hline
        GBDT (E) & 0.49 & 0.96\\\hline
    \end{tabular}
    \normalsize
    \caption{}
    \label{tab:threhsolds}
\end{subfigure}
\caption{
    On the left, the Receiver Operating Characteristic curve (ROC) of the classifiers under analysis, evaluated on the EMBER test set~\cite{anderson2018ember}. 
    The letter inside the parenthesis specifies the dataset used for training the classifier: \emph{E} means EMBER, while \emph{P} implied the use of a larger proprietary dataset.
    The red dashed line highlights the performance of each classifier at 0,1\% False Positive Rate (FPR).
    On the right, the detection thresholds of the classifiers ($\theta$) at 1\% FPR, with the corresponding Detection Rate (DR).
    }
\end{figure}

\myparagraph{Malware Detection Performance.}
\label{sec:exp_roc_malware_classifiers}
Before delving inside the performance of the different attacks against the target classifiers, we first compute the Receiver Operating Characteristic (ROC) of the four models, shown in Figure~\ref{fig:roc}.
The score has been computed on the test set of EMBER v1 dataset~\cite{anderson2018ember}.
For each classifier, we compute the detection threshold $\theta$, and the corresponding Detection Rate (DR) at 0,1\% of False Positive Rate (FPR).
We report these findings in Table~\ref{tab:threhsolds}.
It is clear from Figure~\ref{fig:roc} that the GBDT model outmatches the convolutional networks, which aligns with previously reported results on this dataset~\cite{anderson2018ember}.
This benefit might be connected to the manual feature engineering that is used by the GBDT model, instead of letting the network learn the relevant features itself.
The non-linearity imposed by ReLU activation functions does not seem impact the overall score, implying that the majority of the examples in the dataset can be linearly separated.  However, we do note that at lower false positive rates, the performance 
of the ReLU models does exceed that of the linear activation models, which may support the notion that there are some samples that are difficult to separate and where the non-linearity is useful.  
MalConv shows comparable though somewhat lagging results to both the DNN models trained on EMBER and the GBDT model.  Clearly, the use of a larger and more diverse proprietary dataset does not necessarily improve the generalizability of the end-to-end models, which agrees with previous observations by Coull et al.~\cite{coull2019activation} showing that overfitting may actually be beneficial in malware classification tasks.
For each classifier, we compute the threshold such that the classifier has a $0.1$\% False Positive Rate (FPR).

\subsection{White-box Attacks}
\label{sec:results}

We tested all the differentiable models with the attacks formulated in Section~\ref{sec:practical_manipulations} (\emph{Full DOS}, \emph{Extend} and \emph{Shift}), the header attack~\cite{demetrio2019explaining}, the padding attack~\cite{kolosnjaji2018adversarial}, an iterative implementation of the fast gradient sign method (FGSM) that address both padding and slack space ~\cite{kreuk2018deceiving, suciu2019exploring}. 
Since the \emph{Full DOS} attack searches for the \verb|PE| signature inside the sample, and it marks as editable all the bytes in between the two headers, the amount of bytes to perturb varies from \verb|118| to \verb|290| bytes (since such position it might change from file to file).
We set the length of the adversarial injection for the \emph{Extend} to \verb|512|, and this number is rounded to the nearest multiple of the file alignment specified by the sample: in our test set, it varies between \verb|512| and \verb|4096| bytes, resulting in a payloads whose length varies between \verb|630| and \verb|4386|.
Similarly, we chose for \emph{Shift} attack an injection of \verb|1024| bytes before the first section of the sample, and it must align to the nearest multiple of the specified file alignment, resulting in adversarial payloads with a length between \verb|1024| and \verb|4096| bytes.
The \emph{Partial DOS} attack proposed by Demetrio et al.~\cite{demetrio2019explaining} alters only the first \verb|58| bytes contained in the DOS header.
The \emph{Padding} attack appends bytes at the end of the file, with a default payload size of \verb|10240| KB, motivated by the results obtained by Kolosnjaji et al.~\cite{kolosnjaji2018adversarial}.
The iterative implementation of FGSM has an $\epsilon$ parameter that controls the amount of injected noise, and we set this parameter to $0.1$.
For all the attacks, we have chosen a step size of \verb|256| bytes optimized at each iteration.

\begin{figure}
\begin{subfigure}{.33\textwidth}
    \centering
    \includegraphics[width=\linewidth]{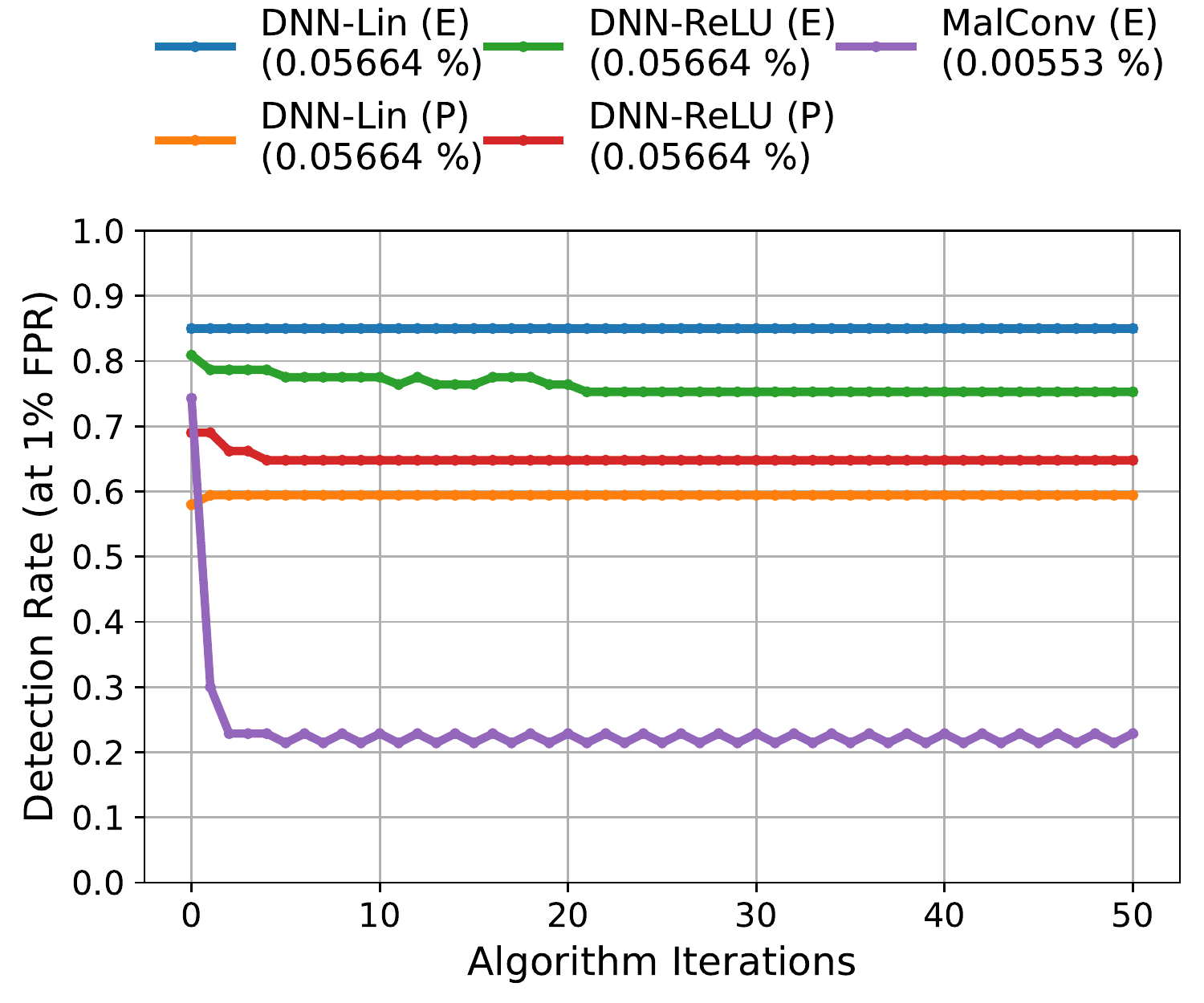}
    \caption{Partial DOS}
    \label{fig:wb_pdos}
\end{subfigure}%
\begin{subfigure}{.33\textwidth}
    \centering
    \includegraphics[width=\linewidth]{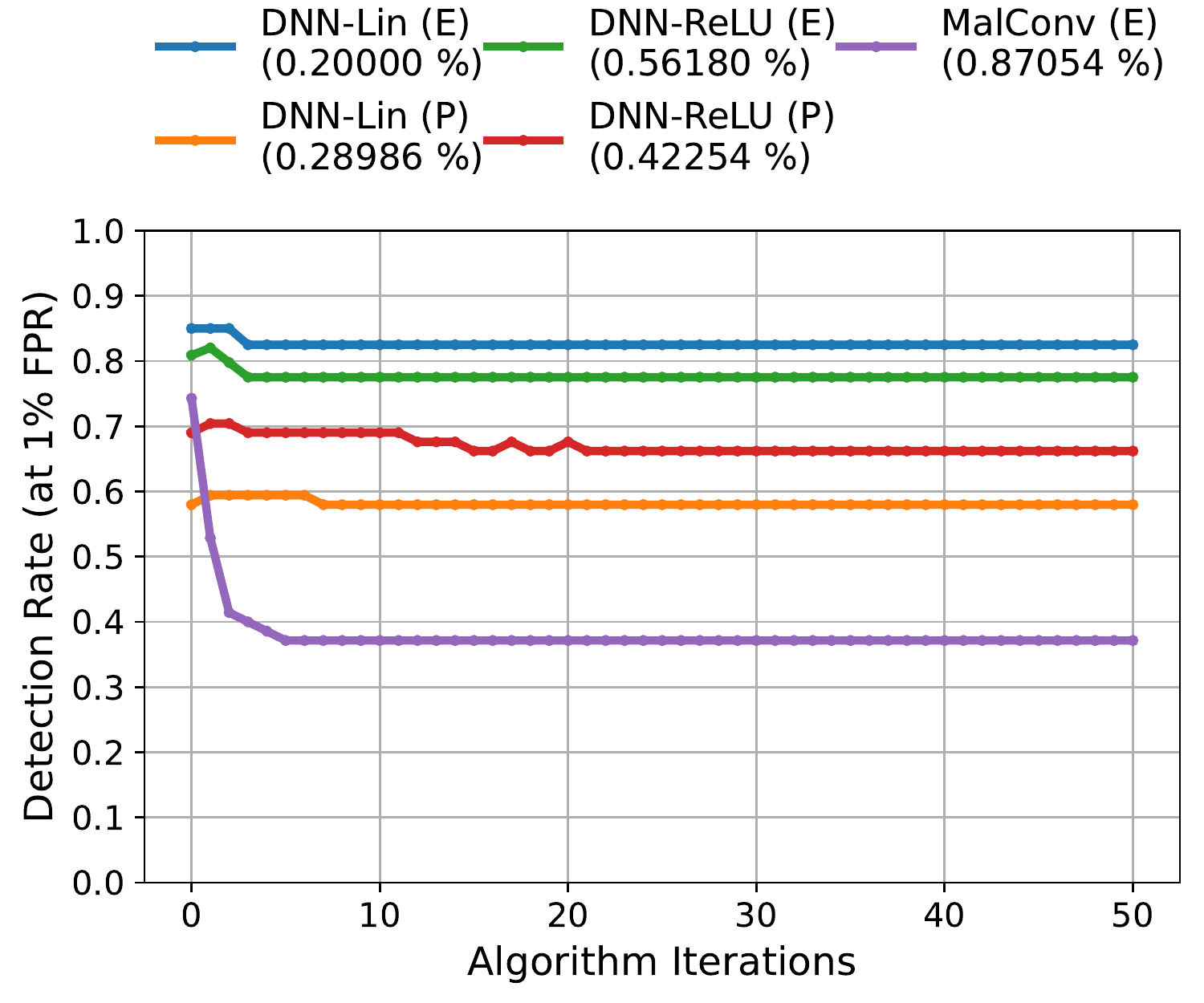}
    \caption{Padding}
    \label{fig:wb_padding}
\end{subfigure}%
\begin{subfigure}{.33\textwidth}
    \centering
    \includegraphics[width=\linewidth]{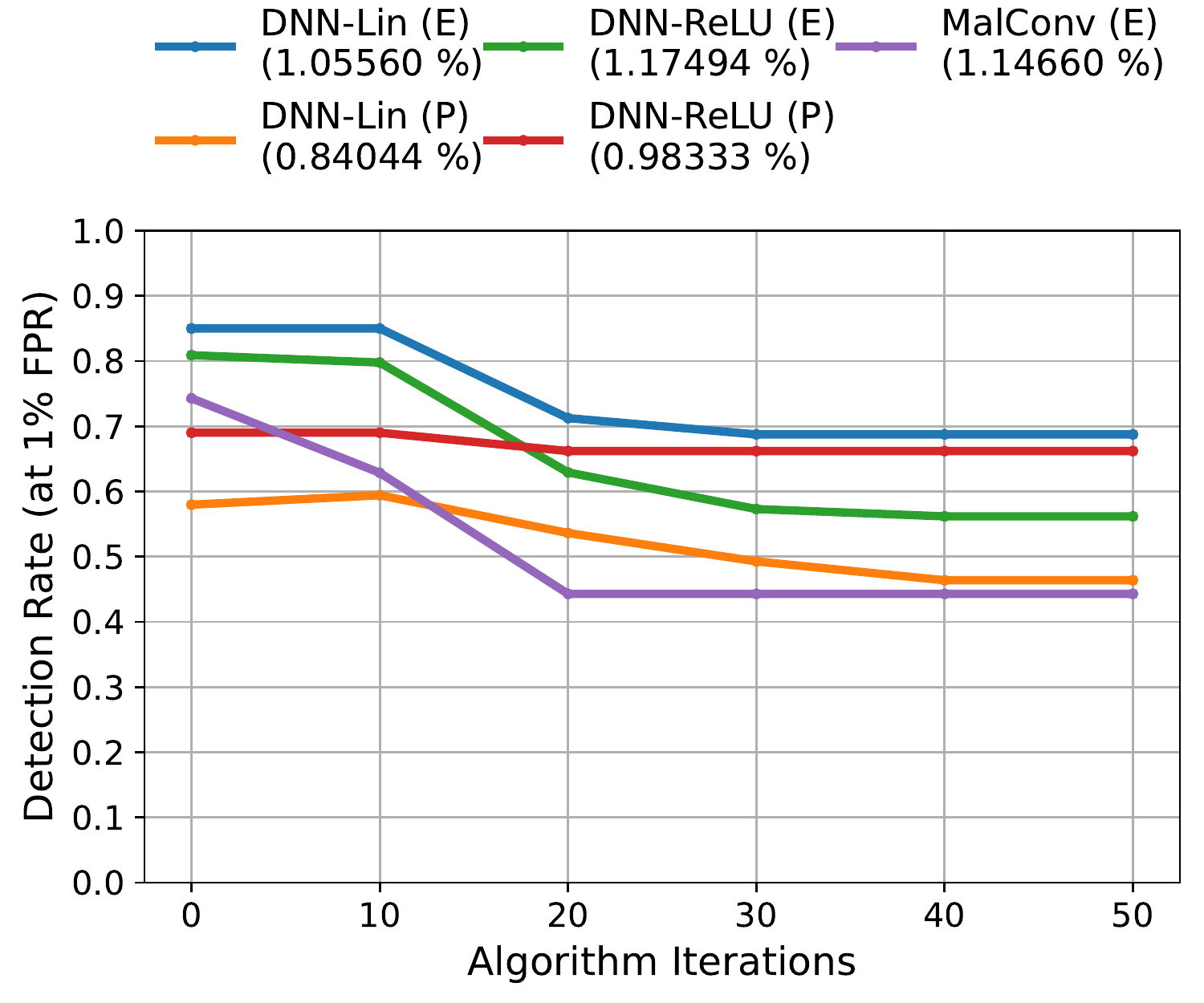}
    \caption{FGSM}
    \label{fig:wb_fgsm}
\end{subfigure}
\hfill
\vspace{3mm}
\begin{subfigure}{.33\textwidth}
    \centering
    \includegraphics[width=\textwidth]{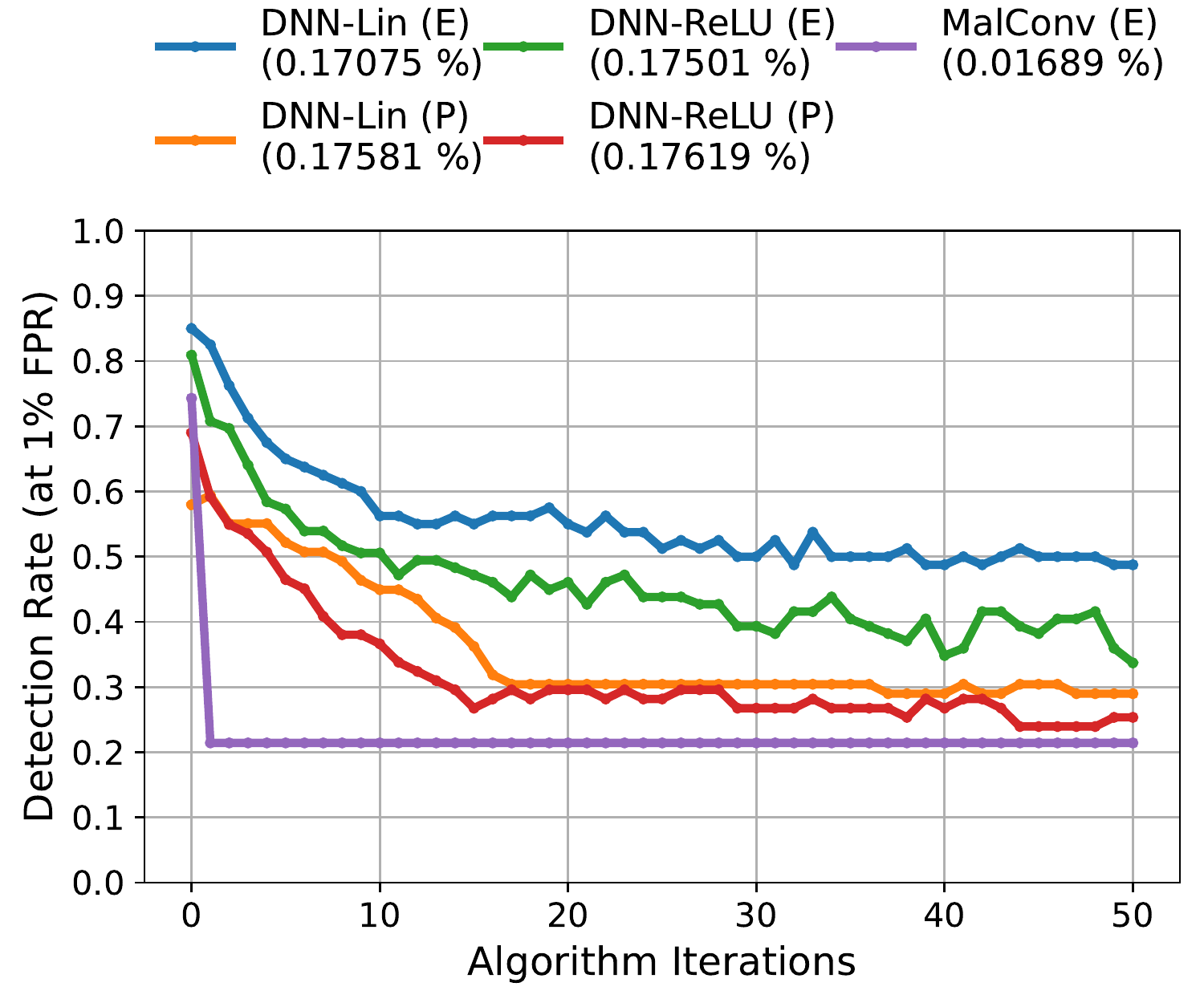}
    \caption{Full DOS}
    \label{fig:wb_fdos}
\end{subfigure}%
\begin{subfigure}{.33\textwidth}
    \centering
    \includegraphics[width=\linewidth]{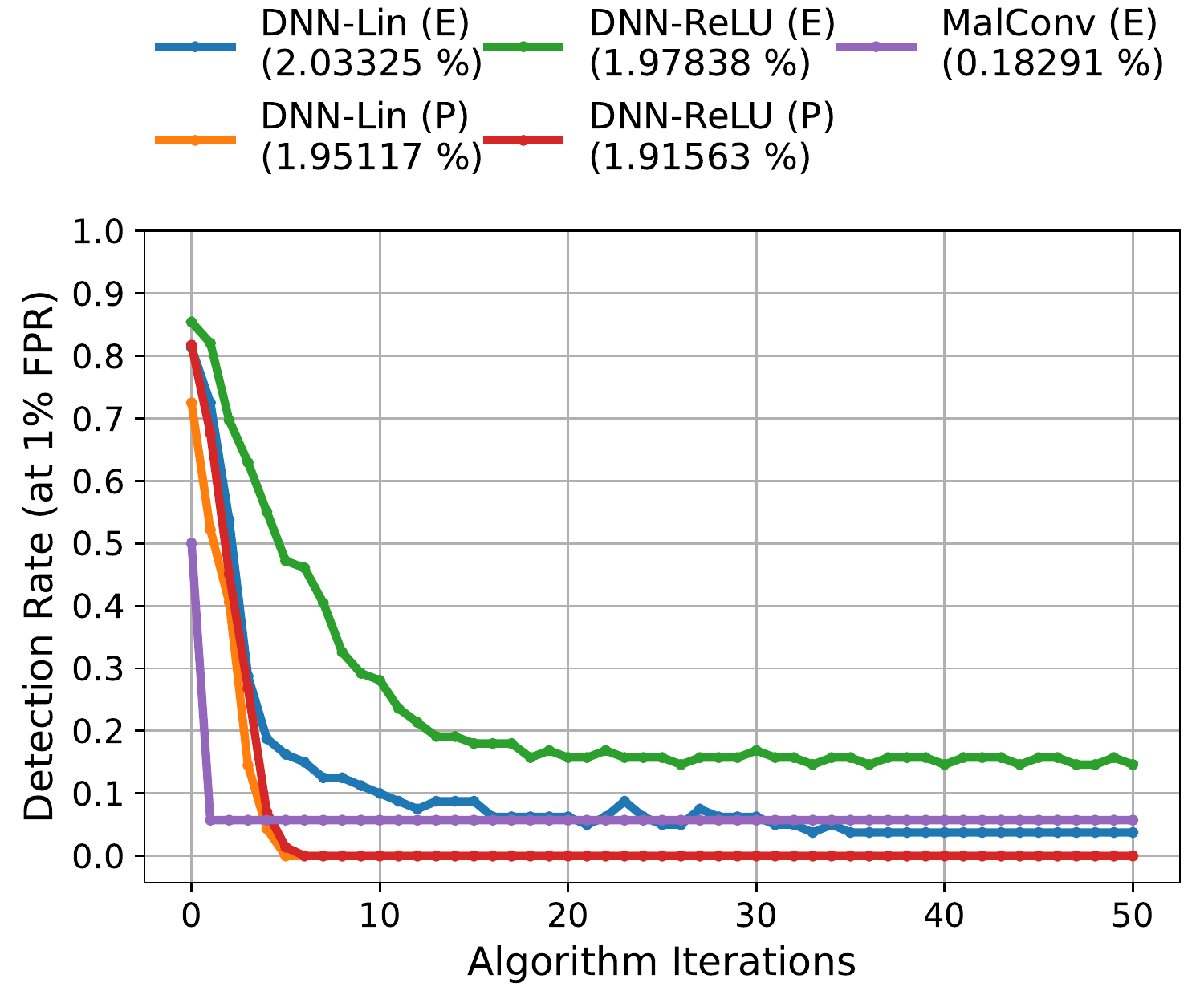}
    \caption{Extend}
    \label{fig:wb_extend}
\end{subfigure}%
\begin{subfigure}{.33\textwidth}
    \centering
    \includegraphics[width=\linewidth]{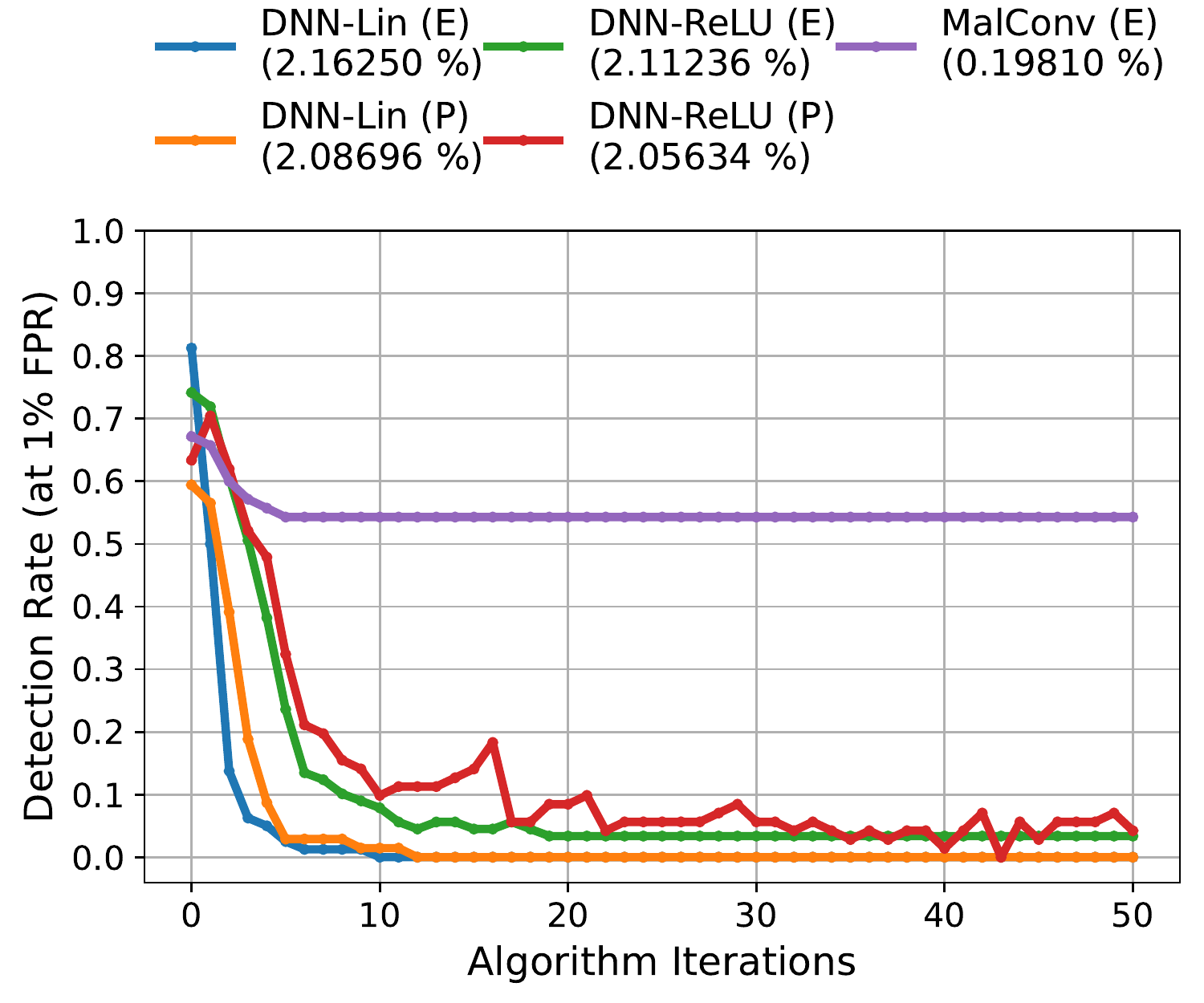}
    \caption{Shift}
    \label{fig:wb_shift}
\end{subfigure}%
\caption{The results of white-box attacks, expressed as the mean Detection Rate at 0.1\% FPR, at each optimization step.
Each plot sums up the degradation induced by a specific strategy against the classifiers we have considered, trained on different datasets (\emph{E} for EMBER and \emph{P} for proprietary).
The number near the name of the classifier represents the size of the adversarial payload as a percentage of the input size.}
\label{fig:fireeye_networks_whitebox}
\end{figure}

While the \emph{Partial DOS} technique is generally ineffective against all classifiers except for MalConv (as already pointed put by Demetrio et al.~\cite{demetrio2019explaining}), the \emph{Full DOS} attack does substantially lower the detection rate of the networks proposed by Coull et al.~\cite{coull2019activation}.
This might be caused by spurious correlations learnt by the network, and altering these values cause the classifier to lose precision.

Both our novel strategies, \emph{Extend} and \emph{Shift} show great attack performance against all evaluated networks.
Since these attacks replace a portion of the real header of the program, it might be possible that the adversarial noise interferes with the local patterns learned by the networks at training time, like the position of the meaningful metadata of the program.
The \emph{Extend} attack, for instance, covers the original position of the \verb|PE| offset plus many fields of the Optional Header, like the \emph{checksum} and the locations of directories such as the Import Table and Export table.
This content is preserved, since it is shifted, but it is no longer present in the position the network believed them to be.
The \emph{Shift} attack does the same, but with the content of the first section, that is usually the one containing the code of the program.
Surprisingly, the \emph{Shift} attack against MalConv is not as effective as it was against the other networks.
The reason might be once again the wrong feature importance that MalConv attributes to certain bytes.
Analyzing the norm of the gradient computed on the location altered by the attack, we found that it is mostly zero, and the attack is unable to optimize the payload.
If the attention is focused on the header, the rest of the file has a low impact on the final score.
This can be glimpsed by looking at the \emph{Extend} attack, which manipulates an extended portion of bytes starting from the DOS header.

The \emph{Padding} attack proposed by Kolosnjaji et al.~\cite{kolosnjaji2018adversarial}, and the \emph{FGSM} attack proposed by Kreuk et al.~\cite{kreuk2018deceiving} and Suciu et al.~\cite{suciu2019exploring} do not decrease much the detection rates of the networks, since most of the manipulations applied are cut off by the limited window size of the network itself.
For instance, if a sample is larger than \verb|100| KB, it can not be padded, and all the strategies that rely on padding fail.
To achieve evasion, these FGSM attacks can only leverage the perturbation of the slack space, but the number of bytes that can be safely manipulated is too few to have significant impact.
Also, this strategy is incapacitated by the inverse-mapping problem: they compute the adversarial examples inside the feature space, and they project them back only at the end of the algorithm.
This means that the attack might be successful inside the feature space, but not inside the input space, where there are a lot of constraints that are ignored by the attack itself.
Against MalConv, the \emph{Padding} attack proves to be quite effective, but it needs at most 10 KB to land successful attacks, as already highlighted by Kolosnjaji et al.~\cite{kolosnjaji2018adversarial}.
The adversarial payload must include as many bytes as possible to counterbalance the high score carried by the ones contained inside the header.

Regarding the size of the adversarial payload injected by our novel attacks, we report the mean percentage size of the crafted noise \wrt to the input window of the target network near every label of the legend of Figure~\ref{fig:fireeye_networks_whitebox}.
This network has a window size of \verb|100 KB|, and each attack alter, on average, 2\% of that quantity (approximately, \verb|3 KB|).
Also, we can observe that both DNN-Lin and DNN-ReLU trained on the larger dataset are less robust \wrt to their counterparts trained on EMBER, and this pattern can be observed in almost every white-box attack we have showed.


\subsection{Black-box Transfer Attacks}
\label{sec:transfer}
We show how the classifiers under analysis behave against black-box transfer attacks.
The latter is crucial since an attacker might optimize attacks against a model they own, and then they can try to evade other systems in the wild.
To this extent, we use the adversarial EXEmples crafted for the white-box attacks, we test them against all the other models, and we use again the threshold $\theta$ at 0,1\% FPR to compute the DR.
\begin{figure}
    \begin{subfigure}{0.33\textwidth}
        \centering
        \includegraphics[width=\linewidth]{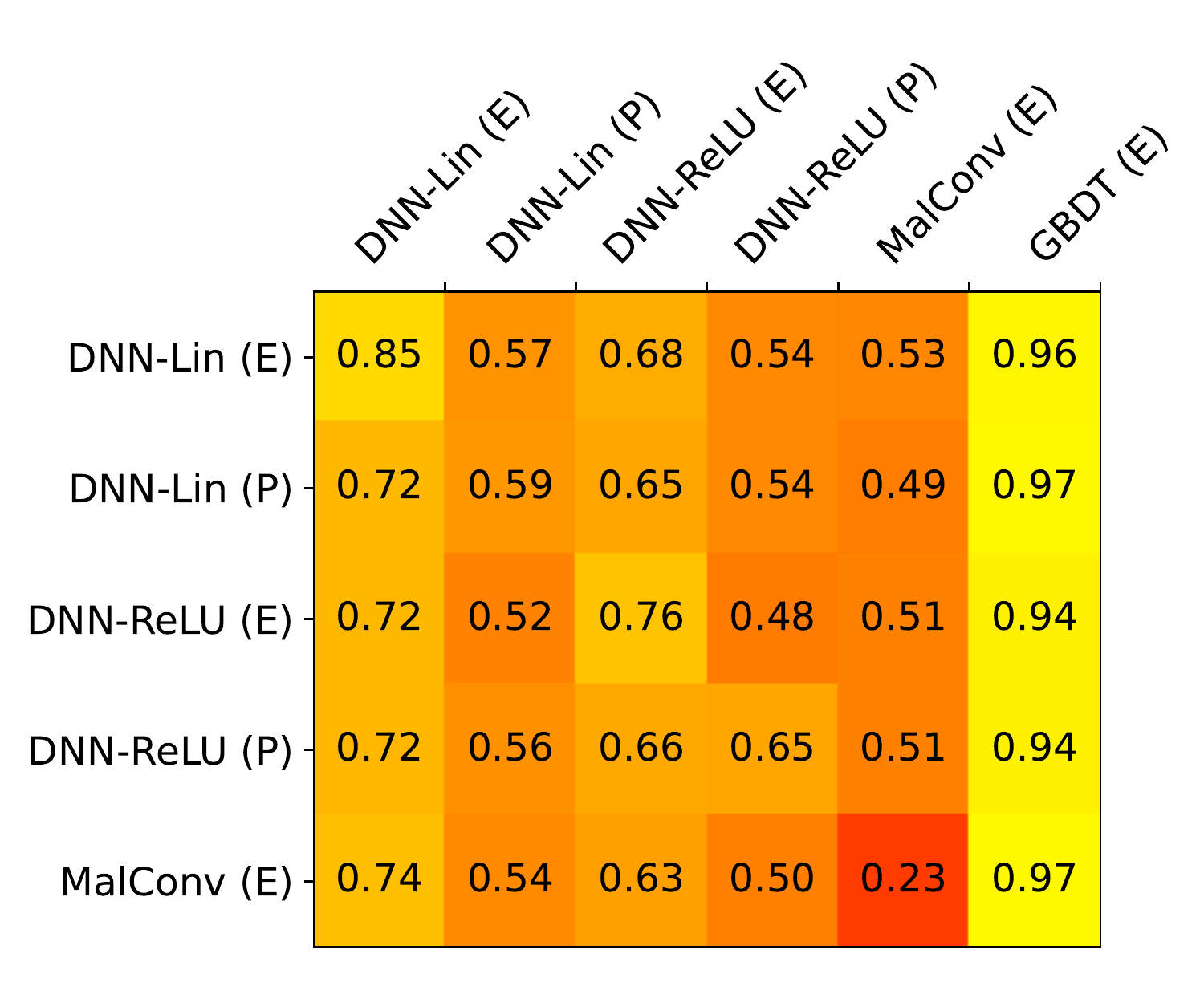}
        \label{fig:transfer_partial_dos}
        \caption{Partial DOS}
    \end{subfigure}
    \begin{subfigure}{0.33\textwidth}
        \centering
        \includegraphics[width=\linewidth]{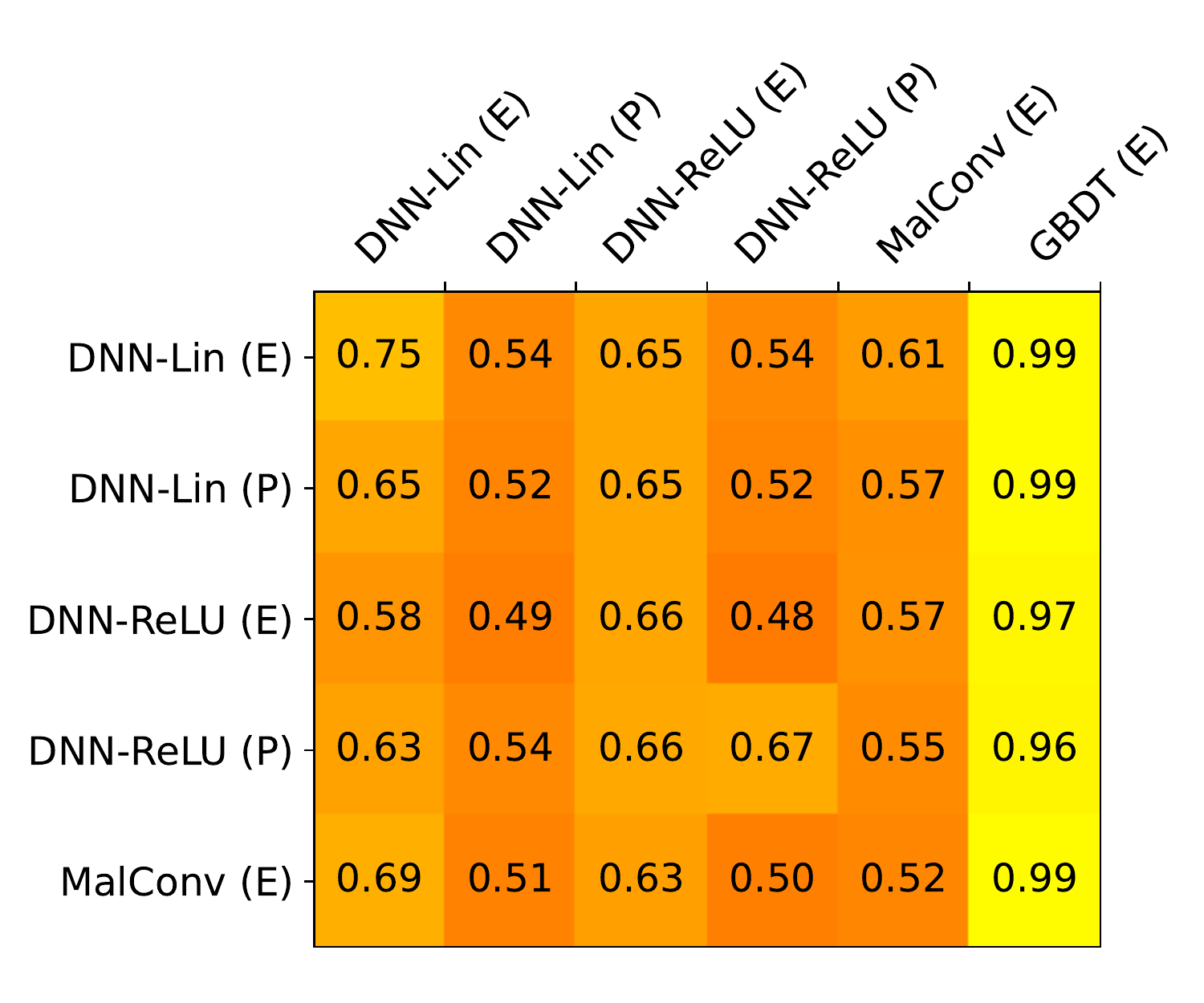}
        \label{fig:transfer_fgsm}
        \caption{FGSM}
    \end{subfigure}
    \begin{subfigure}{0.33\textwidth}
        \centering
        \includegraphics[width=\linewidth]{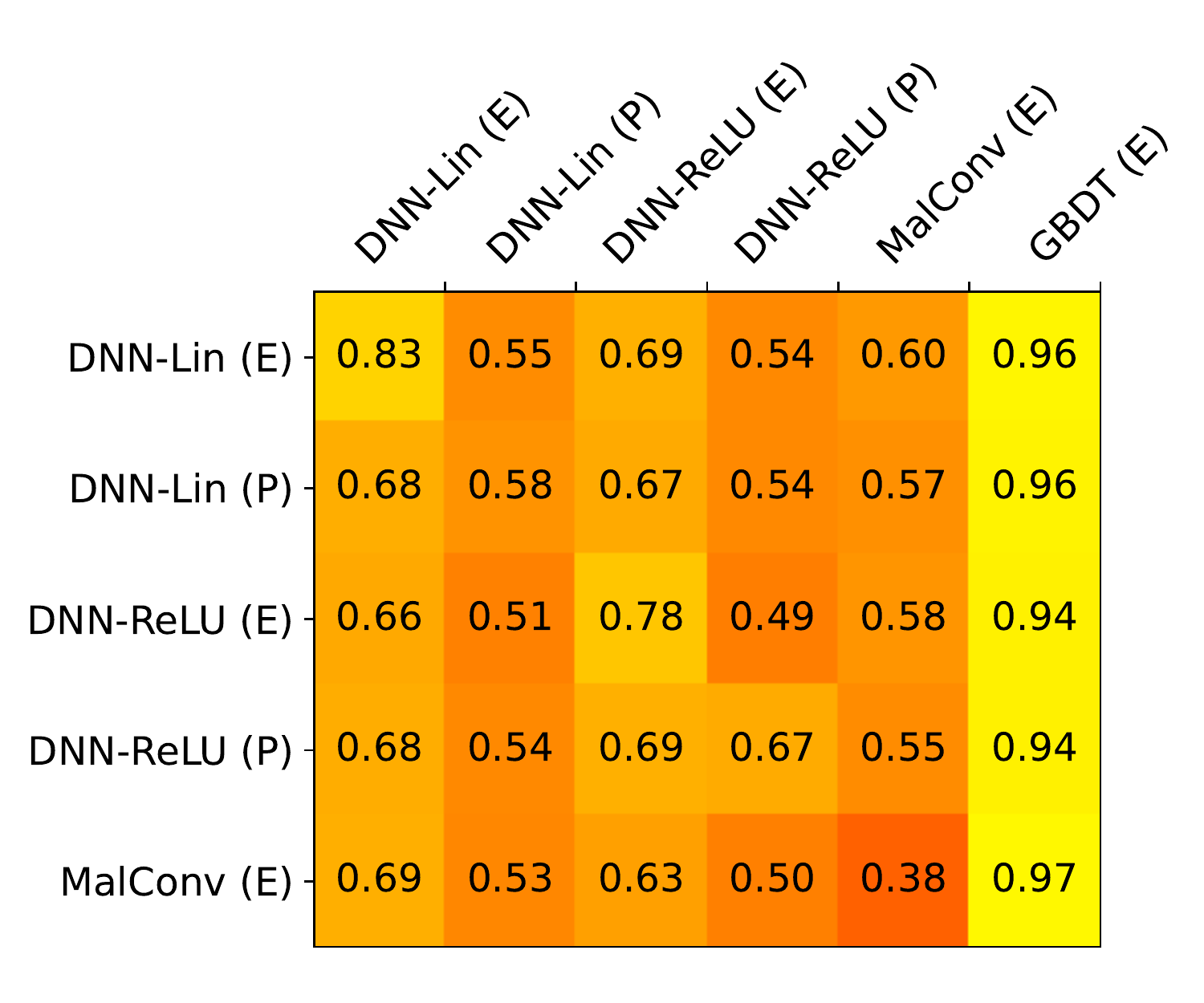}
        \label{fig:transfer_padding}
        \caption{Padding}
    \end{subfigure}
    \hfill
    \begin{subfigure}{0.33\textwidth}
        \centering
        \includegraphics[width=\linewidth]{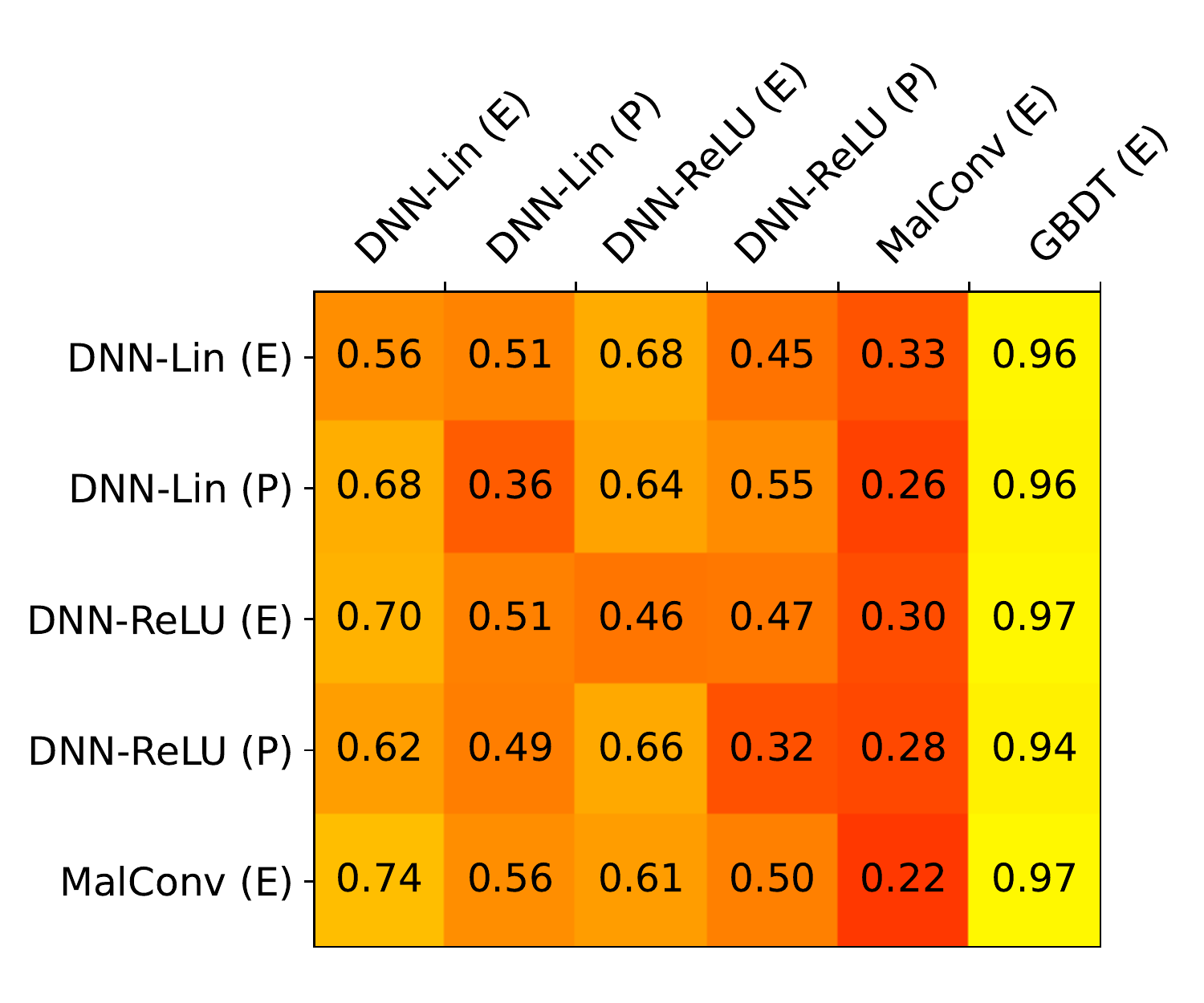}
        \label{fig:transfer_full_dos}
        \caption{Full DOS}
    \end{subfigure}
    \begin{subfigure}{0.33\textwidth}
        \centering
        \includegraphics[width=\linewidth]{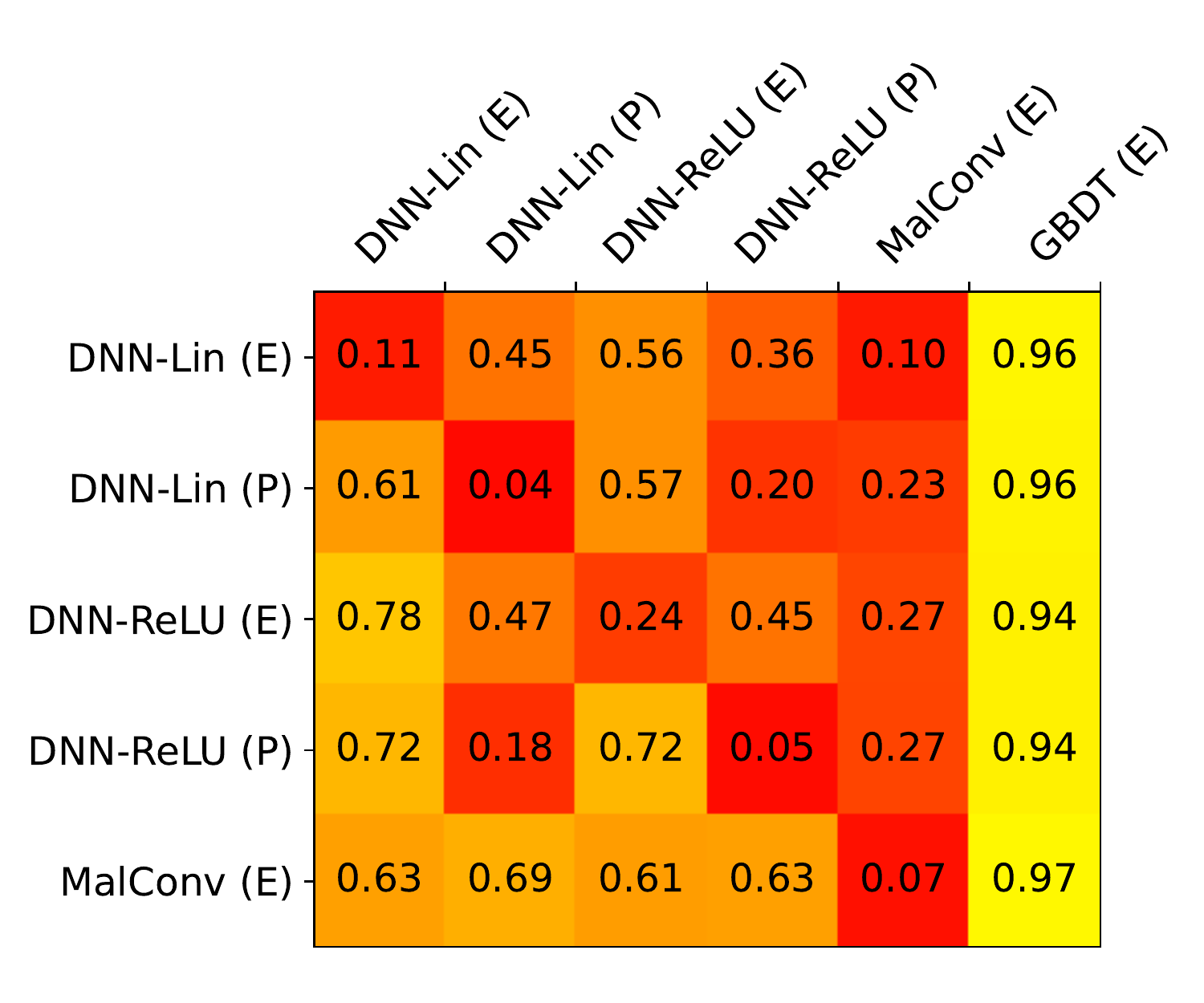}
        \label{fig:transfer_extend}
        \caption{Extend}
    \end{subfigure}
    \begin{subfigure}{0.33\textwidth}
        \centering
        \includegraphics[width=\linewidth]{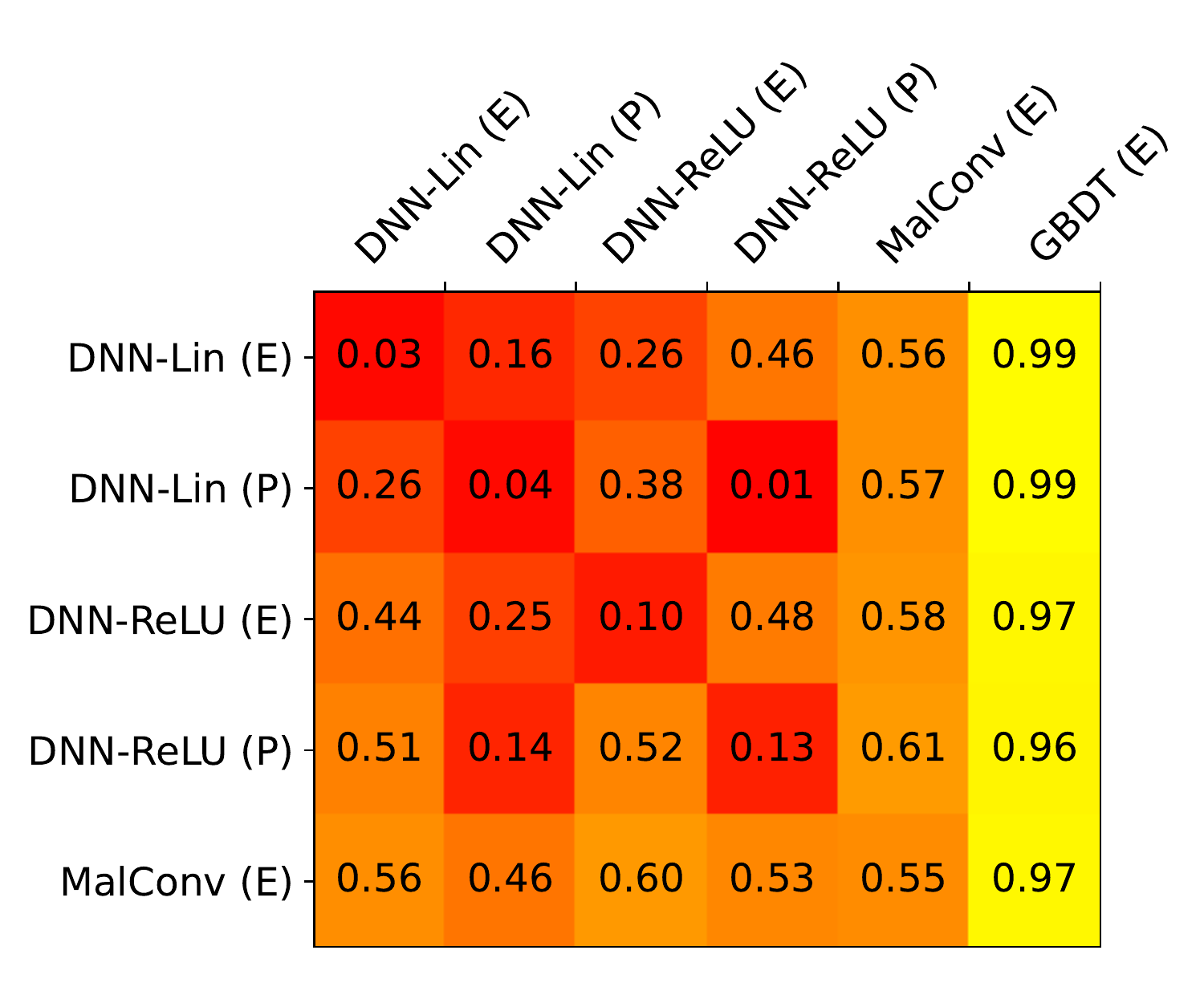}
        \label{fig:transfer_shift}
        \caption{Shift}
    \end{subfigure}
    \caption{Detection rate at 0.1\% FPR for transfer attacks. Substitute models used to craft the attack are reported in rows, while the target models are reported in columns.
    Values on the main diagonal correspond to white-box attacks.
    }
    \label{fig:all_transfer}
\end{figure}
We plot the transfer results on Figure~\ref{fig:all_transfer}, where the rows of each square we report the model we used for computing the attack (\ie, the surrogate), while each columns represent the target of the attack.
Since the GBDT model is not differentiable, we only use it as a target and not as a surrogate.
From various plots in Figure~\ref{fig:all_transfer}, we can highlight different interesting aspects.
The first one is that the new manipulations, \emph{Extend} and \emph{Shift} generally decrease the performance of every byte-based networks.
This phenomenon probably happens because the networks learnt a specific location for particular patterns of bytes (\eg the magic number \texttt{PE} of the PE header), or some section names (\eg \textit{.text}, \textit{.data}, etc.), and association is disrupted by the injection of new content, leading these known values to be placed elsewhere, unable to be find again by the networks.
The \emph{Full DOS} attack has a good impact against MalConv, even in the transfer setting.
Again, this is another empirical proof that MalConv focus on the header of a program, rather than its other components, and it is sufficient to optimize header attacks even against other network to produce successful adversarial EXEmple against it.
The other attacks do not transfer as well as the novel strategies, and they reflect the results obtained in the ROC curve (Figure~\ref{fig:roc}).
The \emph{Partial DOS} transfer attack manipulates too few bytes compared to the other strategies. 
Also, this attack is not optimized directly against the target, so the effect is similar to almost-random noise.
The \emph{FGSM} performs slightly better than the \emph{Padding} attack, since it relies also on the \emph{Slack Space} practical manipulation, that it is less likely to have been cut by the input window of the target model.
Hence, more adversarial noise is considered by the classifier, but not enough for decreasing the detection rate of the targets.
This highlights once again the problems of \emph{Padding} manipulations, since they are easily stripped away.
Another key observation is the similarity of transfer results for the DNN-Lin and DNN-ReLU networks.
Since these networks are similar in architecture and they were trained on the same dataset, it is possible that they also learnt the same blind spots.
The ones trained on the larger proprietary dataset are not more robust than the ones trained on EMBER, but rather they suffer more from transfer attacks.
In the same way, the attacks optimized against them are not very effective against the EMBER versions.
The explanation might consider that the decision boundary learnt over more data is much more complex, filled with more local minimum and maximum.
This phenomenon has already been observed in the past, by Papernot et al.~\cite{papernot2016transferability}, where they show that more complicated models suffer from transfer attacks, caused by the high-non linearity of their decision boundary.
On the other hand, the GBDT model is not affected by any adversarial transfer attack, as the byte-based features used by the decision-tree algorithm are only a small subset of all the characteristics considered by the classifier, such as the API imports, metadata and more.
These attacks are not directly optimized against it, and the quantity of bytes that are altered is very little compared to the whole file size.
For sure, the adversarial payload has a minimal effect on the byte-based features, but not enough to counterbalance all the other ones.

\subsection{Black-box Query Attacks}
\begin{figure}
\begin{subfigure}{.33\textwidth}
    \centering
    \includegraphics[width=\linewidth]{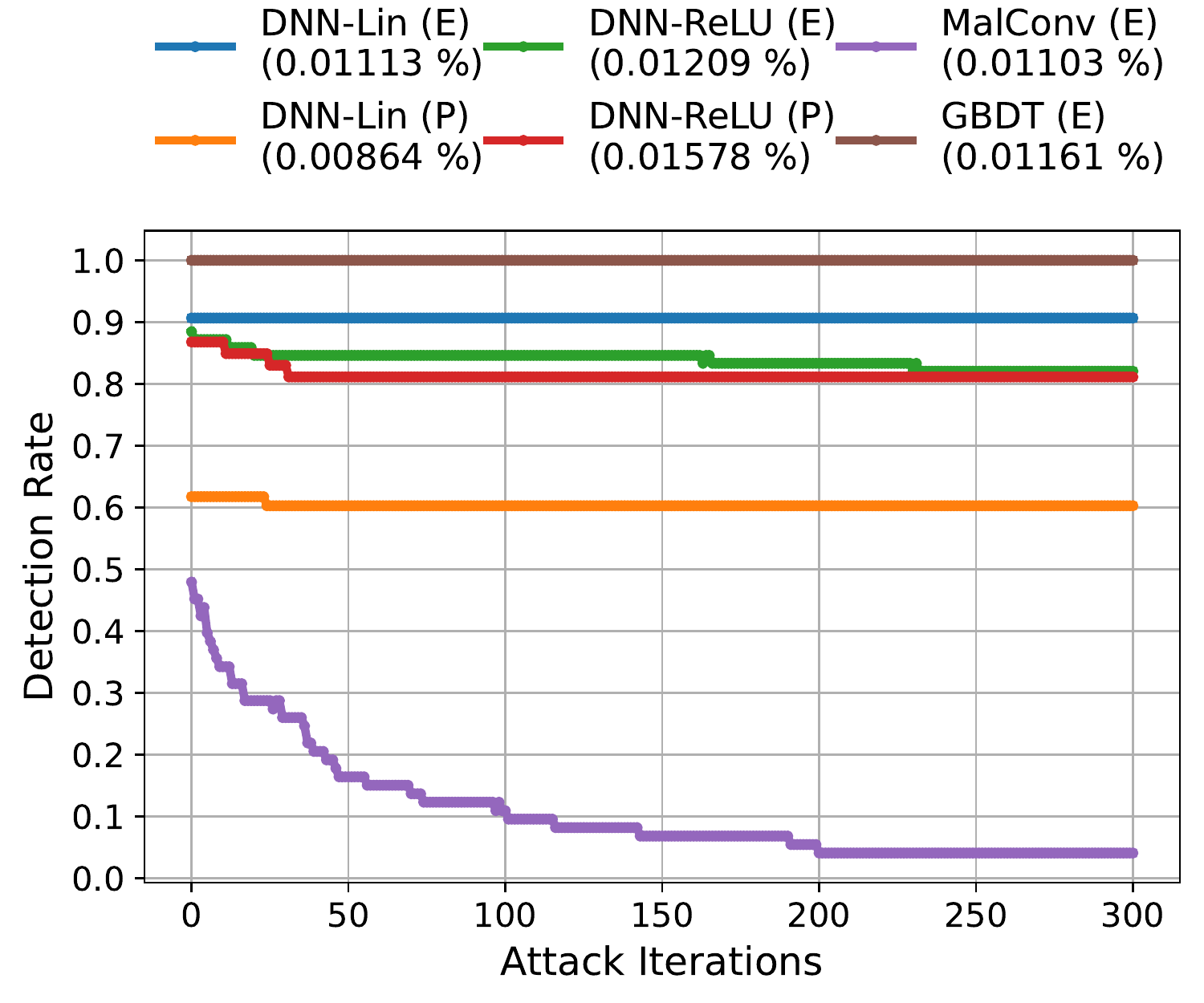}
    \caption{Partial DOS}
    \label{fig:bb_pdos}
\end{subfigure}%
\begin{subfigure}{.33\textwidth}
    \centering
    \includegraphics[width=\linewidth]{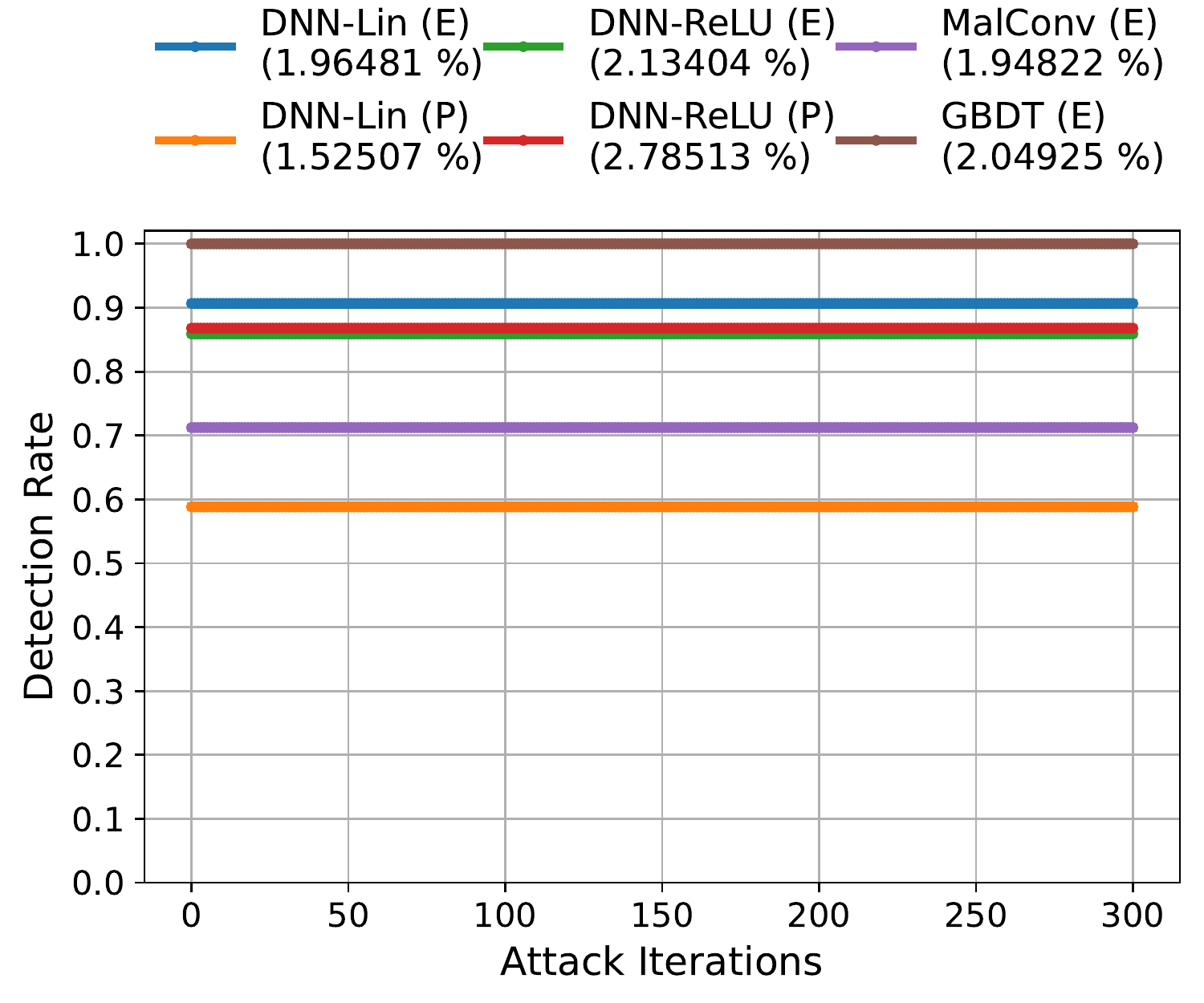}
    \caption{Padding}
    \label{fig:bb_padding}
\end{subfigure}%
\begin{subfigure}{.33\textwidth}
    \centering
    \includegraphics[width=\linewidth]{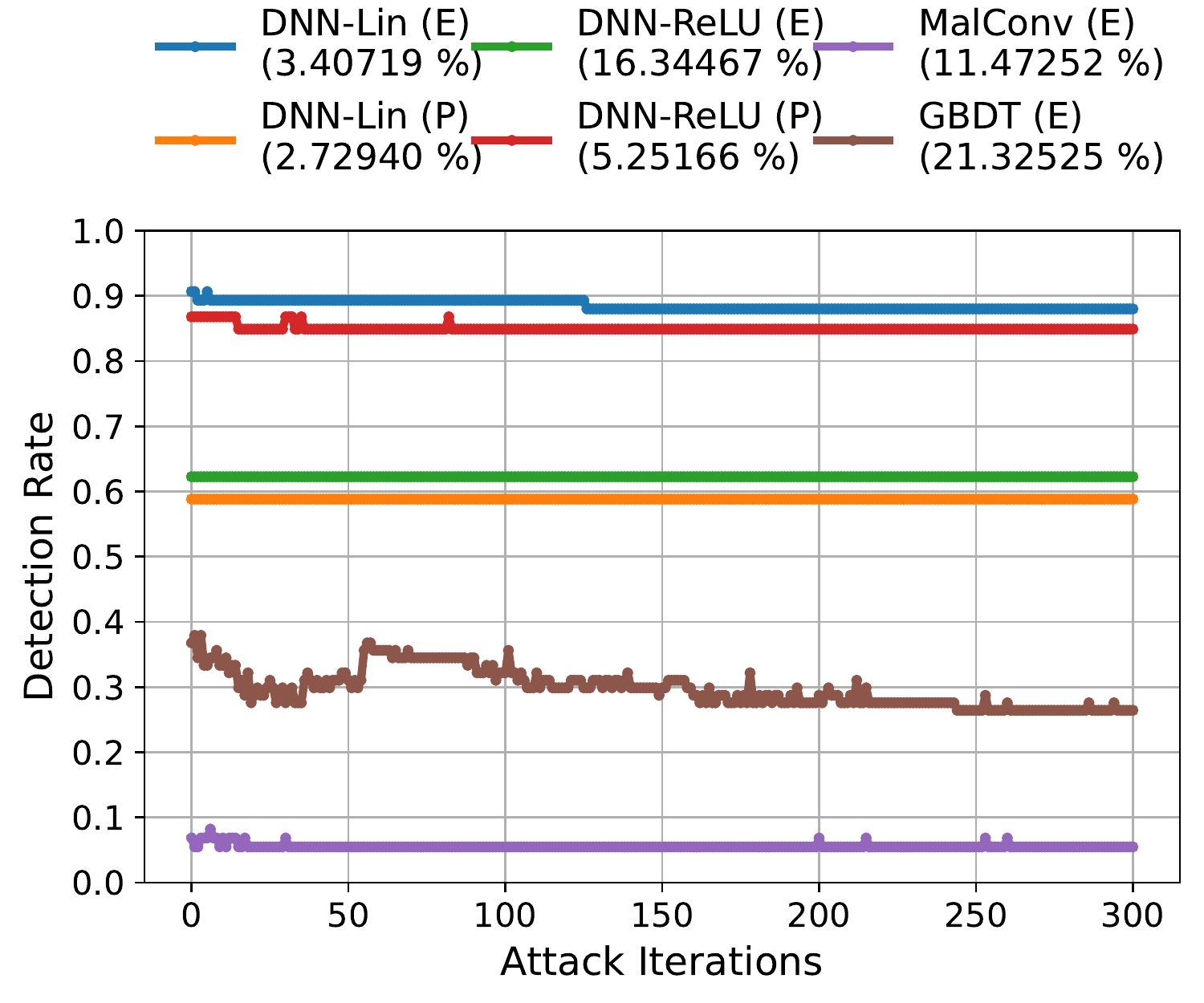}
    \caption{GAMMA}
    \label{fig:bb_gamma}
\end{subfigure}
\hfill
\vspace{3mm}
\begin{subfigure}{.33\textwidth}
    \centering
    \includegraphics[width=\textwidth]{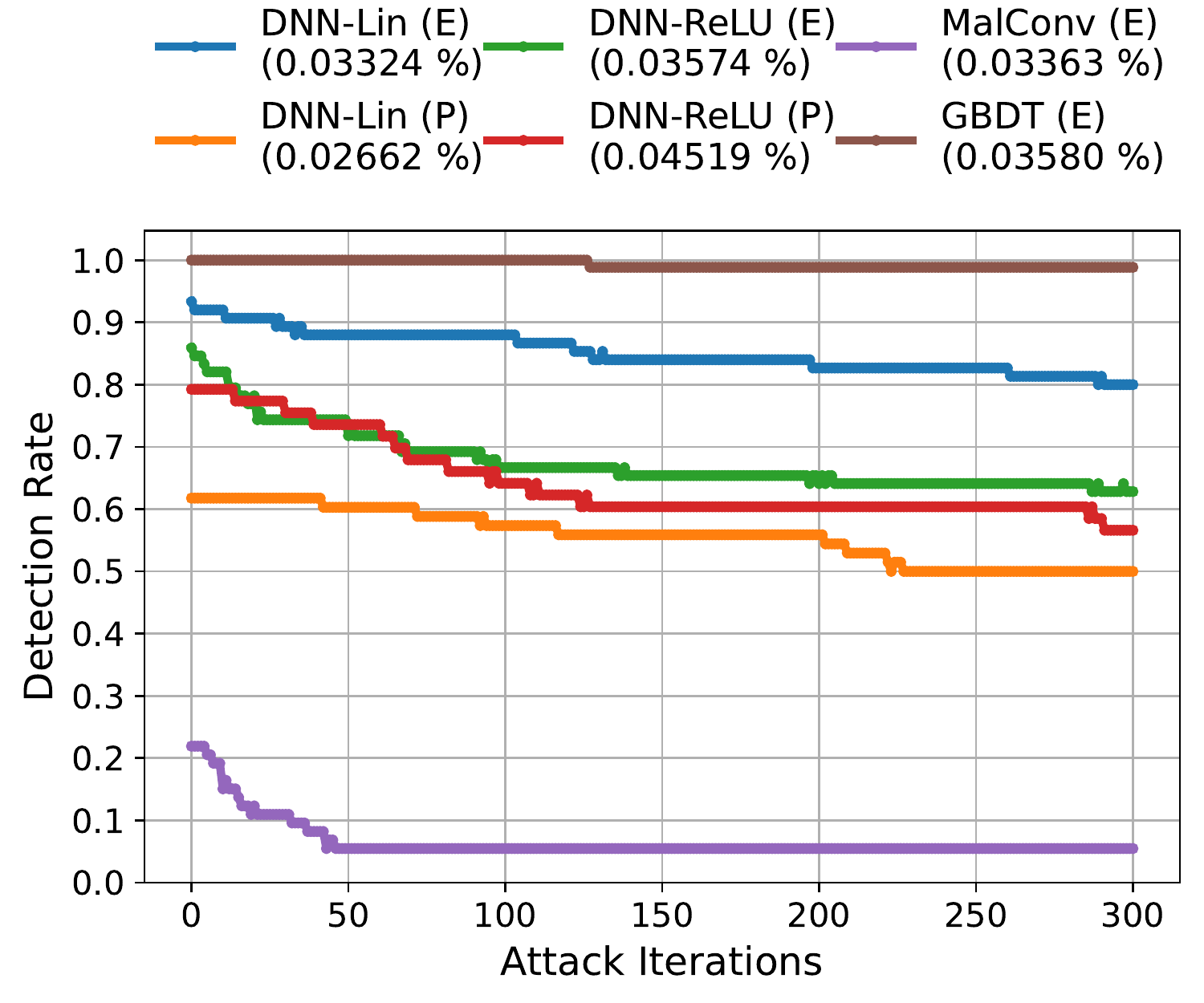}
    \caption{Full DOS}
    \label{fig:bb_fdos}
\end{subfigure}%
\begin{subfigure}{.33\textwidth}
    \centering
    \includegraphics[width=\linewidth]{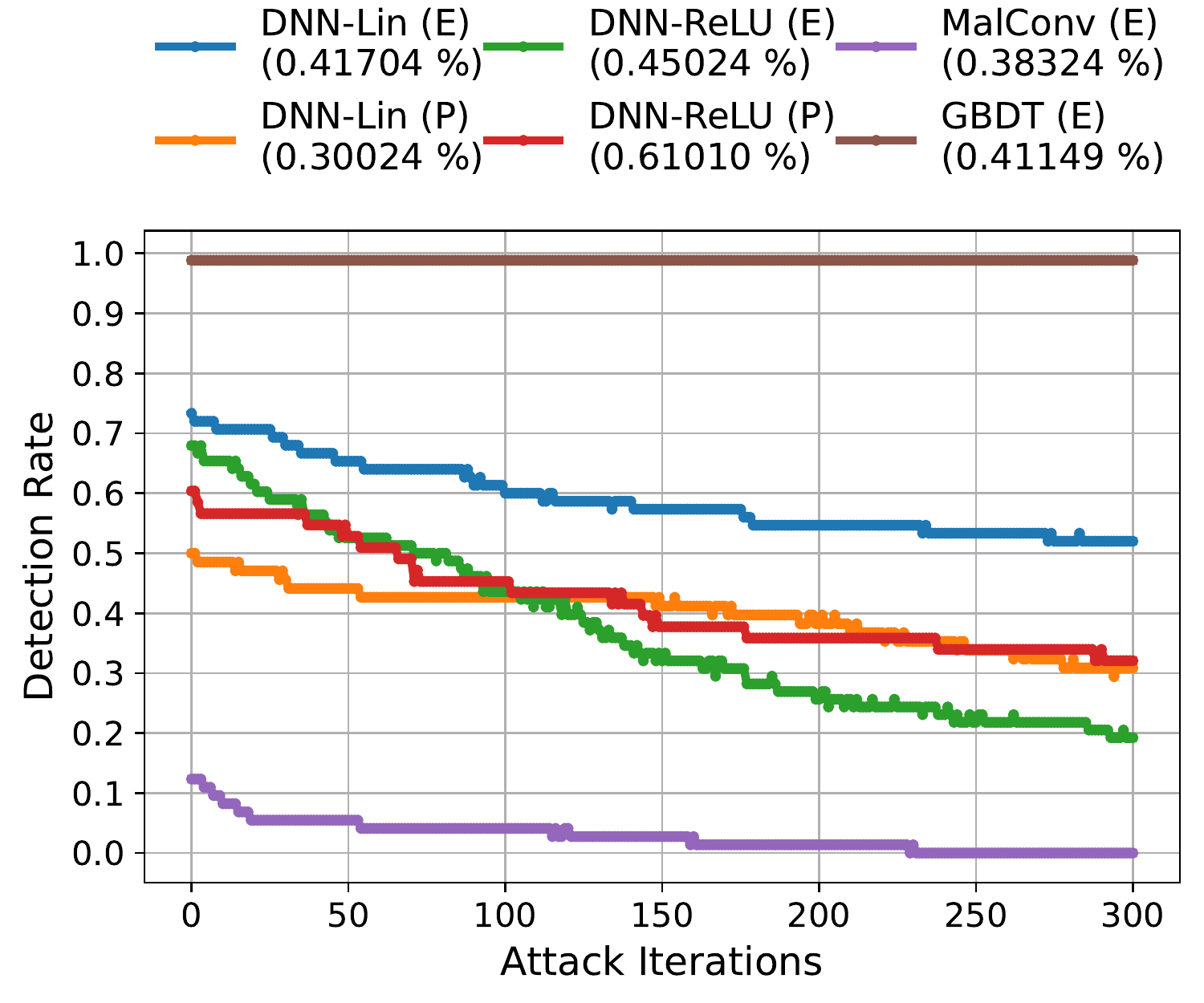}
    \caption{Extend}
    \label{fig:bb_extend}
\end{subfigure}%
\begin{subfigure}{.33\textwidth}
    \centering
    \includegraphics[width=\linewidth]{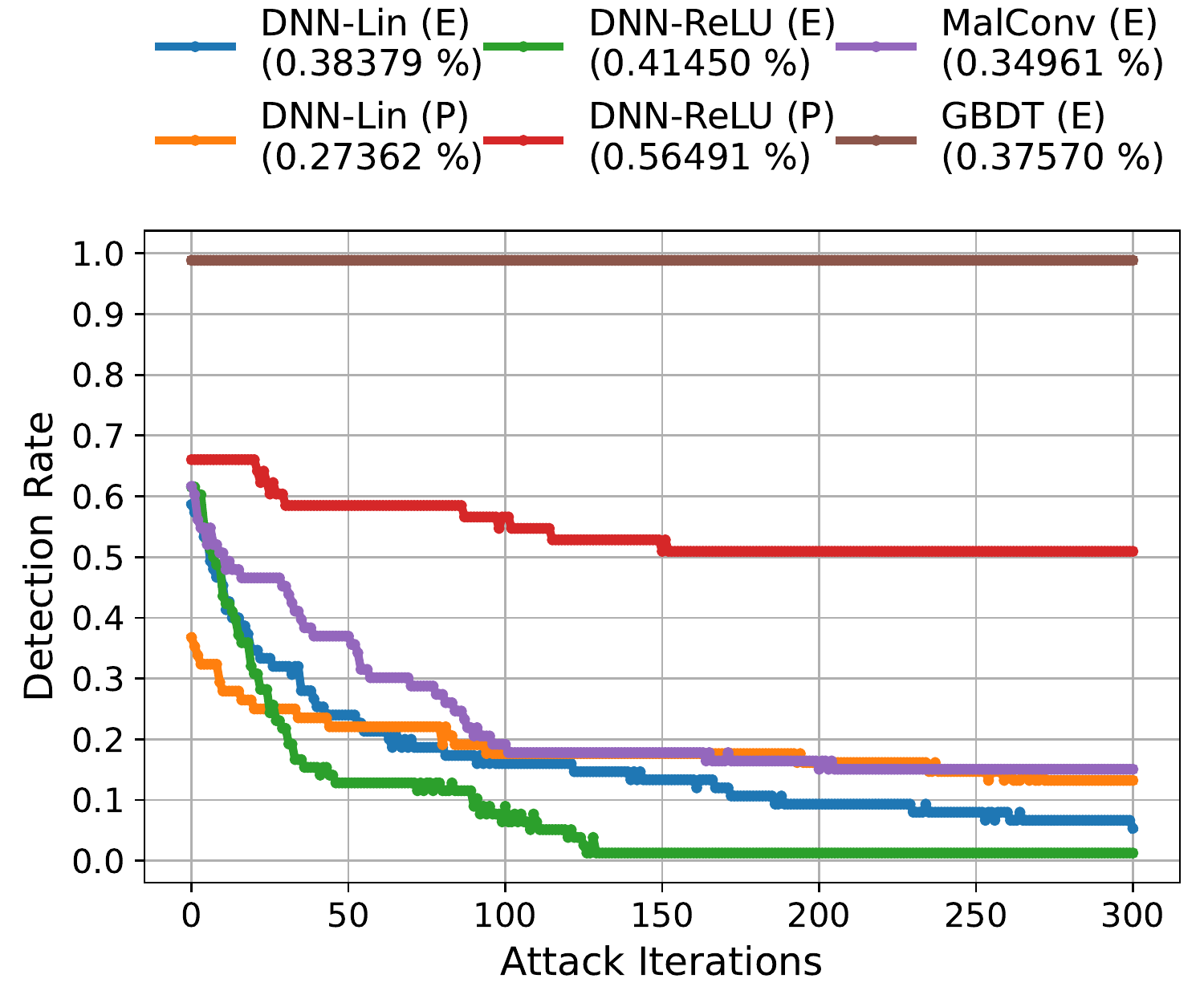}
    \caption{Shift}
    \label{fig:bb_shift}
\end{subfigure}%
\caption{The black-box attacks against all the models. We report the Detection Rate at each step of the black-box optimizer, using a DR at 0.1\% FPR.
The number near the name of the classifier represents the size of the adversarial payload \wrt the mean file size of out malware test set.}
\label{fig:bb_all}
\end{figure}
We evaluate the effects of the black-box query attacks, and we plot the performances of these strategies in Figure~\ref{fig:bb_all}.
We use the mean Detection Rate as metric for this analysis, and we set the population size $N = 10$, and the number of queries to $q=3000$.
This produces $300$ iterations of the genetic algorithm, since each round consumes a number of queries that matches the population size (10 in this case).
Not surprisingly, the first thing we can notice is that the performances of these black-box query attacks are worse \wrt their white-box counterparts.
This is intuitive, since the optimizer does not have any clue regarding a privileged direction to take, and it needs to explore the surrounding space to approximate such direction.
Also, it seems that only the \emph{Shift} attack meaningfully decrease the DR of the networks, while both the \emph{Full DOS} and \emph{Extend} attacks decreases this metric almost linearly in the number of the queries sent.
This might implies that these attacks requires more queries to gain more effectiveness against the byte-based networks.
It is interesting to see that, since these mutations origin from random perturbations, the DNN-Lin and DNN-ReLU trained on the proprietary dataset show more robustness \wrt their counterparts, opposed to the white-box results.
This might be explained by robustness to random noise induced by the high volume of data used for at training time.
Another interesting point, is that MalConv is evaded also by the \emph{Shift} attack, when the same was failing in the white-box settings.
As explained in Section~\ref{sec:results}, the norm of the gradients of the locations perturbed by that manipulations were zero, but the genetic optimizer do not rely on that.
This success is probably caused by the combination of random exploration and optimization done by the optimizer, successfully avoiding regions with no gradient information.
Another important key point is that the GBDT shows great stability against the novel proposed attacks.
This can be caused by a combination of different failures.
We have discussed one of these probable failures while analysing the transfer attacks, in Section~\ref{sec:transfer}: the byte-based features of the GBDT are a minority of all the characteristics that are considered, and these manipulations are not enough for being a relevant contribution during the classification.
Also, the optimizer might not be able to craft patterns that are relevant for the GBDT, since it also has been trained on strings with meaning.
On the other hand, as already proved by Demetrio et al.~\cite{demetrio2020efficient}, the black-box GAMMA algorithm is effective against MalConv and the GBDT.
We considered 100 \emph{.data} sections taken from our goodware dataset, and we use $\lambda = 10^{-5}$, to match the settings of the original formulation. 
The attack, that leverage the \emph{Padding} practical manipulation for injecting benign content, shows similar results to the one obtained by the authors, and it is the only functioning attack against this model.
This is the effect of the injection of benign content: it is difficult for the optimizer to find patterns of bytes without guidance, while using content that already matches the target class is helpful, and the optimizer does not need to create these patterns.
Maybe, by highly increasing the number of queries and the size of the noise, even the other attacks would be successful in the end, but the cost in terms of query budget could be unsustainable.
On the other hand, this strategy is not effective against the DNN-Lin and DNN-ReLU networks, since they focus on a smaller input window than MalConv.
Hence, the adversarial content is stripped away, losing its efficacy.
\section{Related Work}
\label{sec:related}

In this section we briefly review some concurrent work proposing a similar attack framework, provide additional details on attacks against learning-based malware detectors, and conclude by discussing relationships between the problem of adversarial malware to packing and obfuscation techniques.

\subsection{Other Attack Frameworks}
\label{subsec:pierazzi}


Recent concurrent work by Pierazzi et al.~\cite{pierazzi2019intriguing} propose a general formalization for defining the optimization of adversarial attacks inside the input space, spanning multiple domains like image and speech recognition and Android malware detection.
They define sequences of practical manipulations that must preserve the original semantics.
These manipulations need to be imperceptible to manual inspection, and they must be resilient to pre-processing techniques.
The authors also explicitly define the resulting side-effects generated by applying such mutations to the original sample, as a summation of vectors.
The attacker optimizes the sequences of practical manipulations that satisfy all the constraints mentioned above by exploring the space of mutations imposed by such practical manipulations.
Our formalization shares the use of practical manipulations applied inside the input space, and we also minimize the applications of practical manipulations to compute adversarial examples.
Furthermore, our practical manipulations are functionality-preserving by design, including all the constraints proposed by Pierazzi et al, and removing the need of the side-effect vector.
We do not enforce the robustness to the pre-processing step, as the defender needs to know such manipulations in advance to decide what it must be sanitized.
The latter is not trivial, as the novel attacks we propose act as a zero-day, and they can be patched properly only after their discovery.
We generalize the objective function to optimize by including a loss function, and the attacker can choose how to implement it by using a function of their choice, while also adding constraints expressed as regularization parameters.
As opposed to Pierazzi et al., we explicitly express the variables to optimize inside the optimization problem, since they are the parameters of the practical manipulations.

\subsection{Attacks Against Malware Detectors}
Many of the white-box attacks in the space of malware classifiers have been previously introduced in this paper, but we revisit here their respective contributions to the literature here, highlighting how they differ from our work.

Kolosnjaji et al.~\cite{kolosnjaji2016deep} propose an attack against MalConv that leverage padding bytes at the end of the input sample and chooses the best local approximation of each padding value.
Demetrio et al.~\cite{demetrio2019explaining} propose the \emph{Partial DOS} practical manipulation against the MalConv classifier.
Both strategies are easy to apply, since the \emph{Padding} manipulation does not require any particular effort for being applied, but they are not as effective as the novel manipulations we have proposed, as shown in Figure~\ref{fig:fireeye_networks_whitebox} and Figure~\ref{fig:bb_all}, in both white-box and black-box settings.
Also, the number of bytes altered by padding and partial dos are either too few or placed in locations with zero gradient, hence useless during the optimization.

Kreuk et al.~\cite{kreuk2018deceiving} propose an iterative variant of the Fast Gradient Sign Method (FGSM)~\cite{goodfellow2014explaining} by manipulating malware inside the feature space imposed by the target network MalConv using both padding and slack space bytes.
Similarly, Suciu et al.~\cite{suciu2019exploring} apply the classic non-iterative FGSM in feature space by adding bytes to slack space between sections.
Both strategies alter the sample inside the feature space, reconstructing a real adversarial EXEmple only at the end of the iteration, which might reduce the adversarial payload effectiveness.

Sharif et al.~\cite{sharif2019optimization} propose semantics-preserving practical manipulations that alter the code of the input executable, and evaluate against MalConv and the network proposed by Krvcal et al. \cite{krvcal2018deep}.
For instance, they alter math operations, registers, operand and they add instructions without side-effect on the original behavior of the program.
They apply such manipulations at random to each function of the executable, keeping them if the resulting feature vector is aligned with the gradient.
Our approach is entirely guided by the gradient of the target function, and it does not leverage randomness while searching for adversarial examples.
Also, our practical manipulations target the structure of the executable rather than its code, minimizing the size of the perturbation.

\subsection{Malware Detection Through Machine Learning}
We review other techniques that have been produced in the literature to spot malware using machine learning technology.
Saxe et al.~\cite{saxe2015deep} develop a deep neural network which is trained on top of a feature extraction phase. 
The authors consider type-agnostic features, such as imports, bytes and strings distributions along with metadata taken from the headers, for a total of 1024 input variables.
Kolosnjaji et al.~\cite{kolosnjaji2016deep} propose to track which APIs are called by a malware, capturing the execution trace using the Cuckoo sandbox,\footnote{\url{https://cuckoosandbox.org/}} that is a dynamic analysis virtual environment for testing malware.
Hardy et al.~\cite{hardy2016dl4md} statically extract which APIs are called by a program, and they train a deep network over this representation.
David et al.~\cite{david2015deepsign} develop a network that learns new signatures from input malware, by posing the issue as a reconstruction problem. The network infers a new representation of the data, in an end-to-end fashion.
These new signatures can be used as input for other machine learning algorithms.
Incer et al.~\cite{incer2018adversarially} try to tackle the issue of an adversarially-robust classifier by imposing monotonic constraints over the features used for the classification tasks.
Kr{\v{c}}{\'a}l et al.\cite{krvcal2018deep} propose a similar architecture as the one developed by Johns\footnote{\url{https://www.camlis.org/2017/jeffreyjohns}} and Coull et al.~\cite{coull2019activation}: a deep convolutional neural network trained on raw bytes.
Both architectures share a first embedding layer, followed by convolutional layers with ReLU activation functions.
Kr{\v{c}}{\'a}l et al. use of more fully dense connected layers, with Scaled Exponential Linear Units (SeLU)~\cite{klambauer2017self} activation functions.

\subsection{Lessons Learned with Packing}
Another way to achieve evasion without applying any optimization is leveraging packing techniques, \ie a binary rewriting technique that compress a program inside another program, and the latter is decompressed at run-time.
Initially designed to save on disk space and protect intellectual property, packing is often used to obfuscate the representation of input programs.
This causes an increase in the difficulty of studying packed samples, both malign or benign, using static analysis techniques.
In this scenario, machine-learning techniques are not a disruptive technology for detecting threats based only on static information, as described by Aghakhani et al.~\cite{aghakhani2020malware}.
The authors of the work studied how packing decreases the meaningfulness of different static feature sets, by destroying the original representation.
On the other hand, content obfuscation by packing leads to the creation of a new program itself, and it can be seen as a very intrusive way of hiding the malicious content from static analysis-based classifiers.

From the perspective of adversarial machine learning, we are interested in sizing the worst case and evaluate adversarial robustness of machine-learning models by showing that even very small, non-invasive perturbations of the input program can successfully break detection, without the need of packing or obfuscating the whole input program.
The goal of our analyses is to demonstrate how brittle learning-based models can be in face of perturbations carefully optimized against them, rather than showing that static code analysis can be bypassed by packing the whole program.
We do believe that this is really important, as similar issues may also be found for learning-based models trained on features extracted from dynamic code analysis.
In particular,  a learning-based model trained on such features may anyway learn to discriminate between legitimate and malware programs based only a small subset of (very discriminant) feature values. 
In this case, even a small change to such feature values may allowing evading malware detection.
For this reason, we believe that understanding and quantifying adversarial robustness of learning-based malware detectors may not only unveil different, novel vulnerabilities, but that it also constitutes a very important research direction to improve and design more robust models in this space.

\section{Limitations}
\label{sec:limitations}
We discuss here the main limitations of our work, from the efficacy of such methods in other scenarios, to possible mitigation techniques.

\myparagraph{Other Feature Sets.} 
In this work, we showed that manipulations that affect byte-based feature vectors can overthrow end-to-end detectors, by editing only a small portion of the samples.
However, we are aware that these approaches do not impact the performances of detectors that extract information from other objects inside the binary, \eg the imported API, the control flow graph, etc.
Also, as shown by the GBDT~\cite{anderson2018ember}, these manipulations might not be effective against models that partially include byte-based features.
However, the attacker can create suitable practical manipulations for addressing these feature sets, and \algoname can encode them without loss of generality inside the $h$ function (as shown in Section~\ref{sec:practical_manipulations}).
This is the case of the strategy proposed by Demetrio et al.~\cite{demetrio2020efficient}, where they also consider the injection of benign content through the addition of new sections.
This attack strategy is not only able to decrease the performances of the GBDT classifier under 20\% of DR, but it also achieve adversarial evasion against well-known antivirus programs hosted on VirusTotal.~\footnote{\url{https://www.virustotal.com}}
In particular, the authors shows that 12 of them are weak to adversarial attacks, and 9 of them appear in the Gartner Magic Quadrant for Endpoint Protection.~\footnote{\url{https://www.microsoft.com/security/blog/2019/08/23/gartner-names-microsoft-a-leader-in-2019-endpoint-protection-platforms-magic-quadrant/}}

\myparagraph{Dynamic Classifiers.} 
We focused on studying the robustness of classifiers that only address the structure of binary programs.
However, the techniques we have developed would not affect a classifier that consider also the execution of such samples, extracting features like the sequence of system calls or else.
This limitation arises since the content we inject and perturb is not executed at run-time from the program.
The attacker would need to apply \textit{binary re-writing techniques}~\cite{wenzl2019hack}, briefly discussed in Section~\ref{sec:practical_manipulations}, and they are practical manipulations specifically crafted for dealing with dynamic features.
These manipulations leverage the editing of the code of the program, by inserting new branches or swapping instructions that are semantically equivalent, and \algoname is in principle general enough for encoding them.


\myparagraph{Mitigation.} 
Both the \emph{Full DOS} and \emph{Extend} attacks can be easily patched before the classification step, by filtering out all the content between the magic number \verb|MZ| and offset \verb|0x3c|, plus all the content between offset \verb|0x40| and the header identifier \verb|PE|.
Also, the \emph{Shift} attack can be sanitized by looking at whether the beginning of the first section and the end of the optional header match.
The defender can get rid of these manipulations by either erasing the payload or shifting all the content backward and reverting altered section entries.
In both cases, the adversarial payload is deleted, and the classification step can be applied without further complications.
Since we are interested in finding minimal perturbations that alter the decision process, we focused on less invasive alterations, in contrast to the one proposed by Sharif et al.~\cite{sharif2019optimization}, whose application alters the code section of the input program.
However, acknowledging the existence of such manipulations is essential for understanding how to defend against such attacks, and also to understand which feature sets are more useful and stable against adversarial attacks.

\section{Conclusions}
\label{sec:conclusions}
We propose \algoname, a lightweight formalization that encapsulates all the needs of the attacker, with all the practical manipulations applicable in the domain of choice and with all the constraints expressed as a penalty term inside the optimization process.
We define and apply new practical manipulations, namely \emph{Full DOS}, \emph{Extend} and \emph{Shift}, crafted for the domain of malware detection of Windows PE programs, and, we offer an overview of all the practical manipulations that address such domain.
We show hot to recast all the white-box and black-box attacks proposed in the \SoA literature inside \algoname without loss of generality, and implemented accordingly.
We conduct experiments for assessing the results of our new techniques, taking into account state-of-the-art deep learning classifiers, and we produce evasive adversarial malware against them, in both gradient-based (white-box) and gradient-free (black-box) settings.
The amount of noise added to the original malware samples is below 2\% of the input size of the target network, showing that the attacker does not need to rely on complex manipulations to achieve evasion, but rather apply only a minimal perturbation.
The novel white-box attacks shows that the newly-proposed \emph{Extend} and \emph{Shift} attacks are the most effective ones in our experimental analysis.
We test the performance of transfer attacks, showing how an attacker can take advantage of only using surrogate models they own, and we highlight how the networks trained on a larger dataset are weak to these strategies.
Lastly, we show the attacker is able to decrease the detection rate of remote classifiers, by only sending queries and optimizing the adversarial noise based on such responses.

\myparagraph{Future work.} We believe that future lines of research should include the study of which feature sets are easier to perturb with practical manipulations to craft adversarial malware and which are not.
Since semantics have multiple syntactical representations, it can hardly be inferred by static features, as syntax can be easily twisted to shape adversarial malware that evades detection, hence it is interesting to have an hint on the possible moves of a possible attacker.
The formalization we propose is general enough for including also attacks against dynamic and hybrid detectors, since \algoname is highly modular, and it can comply with the attacker's needs.
Also, as previously mentioned, it would be indeed useful to encode more manipulations in this formalization, ranging from perturbations of more complex feature sets to dynamic features.
On a different note, it would be interesting to improve the robustness of machine learning malware classifiers by leveraging domain knowledge in the form of constraints and regularizers, \eg, to capture invariant transformations known to domain experts that may modify the input program bytes but preserve its semantics and functionality~\cite{melacci2020domain}.
This would enable learning robust algorithms against such transformations in a very efficient manner, without the need of performing \emph{adversarial training}, \ie, generating the actual adversarial EXEmples and retrain the model using them. 
In fact, such adversarial training procedure may anyway remain ineffective due to the high number of dimensions of the input space and variability of  the transformations. 
For this reason, we believe that encoding domain knowledge directly into the learning process may substantially improve model robustness by bridging the gap between models that are learned entirely in a data-driven manner and the design of hand-crafted feature representations for them.

\section*{Acknowledgement}
This work was partly supported by the PRIN 2017 project RexLearn (grant no. 2017TWNMH2), funded by the Italian Ministry of Education, University and Research; and by the EU H2020 project ALOHA, under the European Union’s Horizon 2020 research and innovation programme (grant no. 780788).

\bibliographystyle{abbrv} 
\bibliography{bibliography}

\end{document}